\documentclass[useAMS,usenatbib]{mn2e}

\usepackage{amsmath}
\usepackage{amssymb}
\usepackage{graphicx}
\usepackage{multirow}
\usepackage[font=small,labelfont=bf,labelsep=period]{caption}

\title[]
{Why stellar feedback promotes disc formation in simulated galaxies}
\author[]
{Hannah \"Ubler$^{1}$\thanks{E-mail: hannah@mpa-garching.mpg.de},
Thorsten Naab$^{1}$,
Ludwig Oser$^{2}$,
Michael Aumer$^{1}$,
Laura V. Sales$^{3}$,
\newauthor
Simon D. M. White$^{1}$\\
$^{1}$Max-Planck-Institut f\"ur Astrophysik, 85741 Garching, Germany\\
$^{2}$Department of Astronomy and Astrophysics, Columbia University, New York, NY 10027, USA\\
$^{3}$Harvard-Smithsonian Center for Astrophysics, Cambridge, MA 02138, USA\\
}
\date{Accepted 2014 June 21.  Received 2014 June 20; in original form 2014 March 24}

\pagerange{\pageref{firstpage}--\pageref{lastpage}} \pubyear{2014}

\begin{document}
\label{firstpage}
\maketitle

\begin{abstract}
We study how feedback influences baryon infall onto galaxies using
cosmological, zoom-in simulations of haloes with present mass
$\mathrm{M}_{\mathrm{vir}}=6.9\times10^{11} \mathrm{M}_{\odot}$ to
$1.7\times10^{12} \mathrm{M}_{\odot}$. Starting at $z=4$ from
identical initial conditions, implementations of weak and strong
stellar feedback produce bulge- and disc-dominated galaxies,
respectively.  Strong feedback favours disc formation: (1) because
conversion of gas into stars is suppressed at early times, as required
by abundance matching arguments, resulting in flat star formation
histories and higher gas fractions; (2) because $50\%$ of the stars form {\it
in situ} from recycled disc gas with angular momentum only weakly related to
that of the $z=0$ dark halo; (3) because late-time gas accretion is
typically an order of magnitude stronger and has higher specific
angular momentum, with recycled gas dominating over primordial infall;
(4) because 25--30\% of the total accreted gas is ejected entirely before $z\sim1$, removing primarily low angular momentum material which
enriches the nearby inter-galactic medium. Most recycled gas roughly
conserves its angular momentum, but material ejected for long times
and to large radii can gain significant angular momentum before
re-accretion.  These processes lower galaxy formation efficiency in
addition to promoting disc formation.\\
\end{abstract}


\section{Introduction}
\label{intro}

In current cosmological models the formation of galaxies is connected
to hierarchical clustering. Small structures form first, grow, and
merge into larger objects. In this framework, galaxies form through
the cooling of gas at the centers of dark matter haloes where it
condenses into stars. To match the observed 
properties of galaxies and galaxy clusters, purely gravitational processes on their own cannot account
for cosmological structure formation but gas dissipation processes
have to be considered. Therefore it was
suggested early on \citep{1978MNRAS.183..341W} that at high redshift
gas has to be prevented from excessive cooling into dense regions
possibly by feedback from massive stars \citep{1974MNRAS.169..229L, 1986ApJ...303...39D, 1991ApJ...380..320N}. Mechanisms like gaseous galactic outflows were proposed to remove potentially star forming low angular momentum
material at early times during the formation of galaxies
\citep[e.g.][]{2001MNRAS.321..471B}. Direct observational evidence of
the last decades underlines  potential impact of feedback events
\citep[e.g.][]{1999ApJ...513..156M, 2001MNRAS.321..450T, 2003RMxAC..17...47H,
  2006ApJ...637..648S, 2009ApJS..181..272G, 2010ApJ...717..289S, 2011ApJ...733L..16S}. 

Numerical simulations of galaxy formation are an important tool
for understanding the impact of feedback. In simulations, the stellar
components of galaxies are usually made from cooling halo gas and
from gas and stars added by mergers. During gas poor stellar
mergers the orbitals of stars are scrambled,
the stellar systems become dynamically hot, and a significant part of
the stellar angular momentum is transported outwards to the dark
matter component \citep[e.g.][]{1992ApJ...393..484B, 1992ARA&A..30..705B}.
To produce dynamically cold and thin stellar discs, the accretion of
high angular momentum gas from outer regions of the haloes is needed in
the more recent past \citep{1979Natur.281..200F}. This calls for gas
reservoirs at low redshifts as well as for feedback processes at
higher redshifts to avoid early over-cooling and overly efficient star
formation.

Some progress has been made in simulating galaxies by increasing
resolution \citep{2003ApJ...597...21A, 2004ApJ...607..688G} or applying more elaborate 
models for the interstellar medium and star formation
\citep[e.g.][]{2008ApJ...680.1083R, 2010Natur.463..203G, 2012MNRAS.425.3058C,
  2012MNRAS.421.3488H, 2013MNRAS.432.2647H} or for stellar feedback
\citep[e.g.][]{1999ApJ...519..501S, 2001ApJ...555L..17T,
  2005MNRAS.363.1299O, 2008MNRAS.389.1137S, 2011MNRAS.410.1391A,
  2012MNRAS.426..140D, 2013MNRAS.428..129S, 2013ApJ...777L..38H, 2013MNRAS.434.3142A,
  2013MNRAS.436.3031V}. Commonly known problems in disc galaxy
formation such as undersized disc galaxies and the significant loss of
angular momentum from gas particles to the dark matter component,
known as the angular momentum catastrophe \cite[e.g.][]{1991ApJ...380..320N, 1995MNRAS.275...56N, 1998MNRAS.300..773W, 2000ApJ...538..477N, 2002MNRAS.335..487M}, have been worked on for
several years. By now, many of the past problems have been solved and
more veritable disc galaxies can be formed
\citep[e.g.][]{2009MNRAS.396..696S, 2011ApJ...742...76G,
  2011MNRAS.410.2625P, 2011MNRAS.410.1391A, 2012MNRAS.424.1275B, 2013MNRAS.434.3142A,
  2014MNRAS.437.1750M}. 

In particular the study of the evolution of the gas component in
simulations will help to better understand the physical
processes regulating galaxy formation. \cite{2008MNRAS.387..577O} and
\cite{2010MNRAS.406.2325O} introduced the concept of `wind recycling',
describing fluid elements which are ejected in a wind, then
re-accreted and ejected again or, alternatively, condensed completely
into stars. They tracked the gas particles during
hydrodynamical SPH simulations and monitored if and how often they
entered a wind mode. Most `wind' particles do not stay in the
intergalactic medium (IGM) but are re-accreted
onto the galactic halo possibly several times. They may never actually reach the IGM but circle within a so called `halo fountain', the name chosen
as reference to the `galactic fountain', first
  introduced by \cite{1976ApJ...205..762S}, which describes a similar 
process on smaller galactic scales. The halo fountain is the
dominating recycling process at late times in the simulations of Oppenheimer et al.
($z \leq 1$), when the 
wind particles typically remain within the parent halo. 

\cite{2011MNRAS.415.1051B} found in their simulations with
supernova-powered outflows that during the assembly of the galaxy, low
angular momentum material tends to be ejected at early times when the
potential wells of the forming galaxies are still shallow. In
addition, gas is primarily ejected perpendicular to the disc whereas
inflow occurs mainly in the disc plane \citep[but see][]{1309.5951}. In a
subsequent paper \cite{2012MNRAS.419..771B} found that the ejected gas is sometimes re-accreted at later times with additional angular momentum
gained through mixing with hot corona gas. While the ejection of
gas was found to be an important process at all galaxy mass scales,
the redistribution of angular momentum via the re-accretion of gas in
galactic fountains gets more relevant for higher mass
galaxies. \citeauthor{2012MNRAS.419..771B} therefore concluded that galactic fountains may lead to the formation of
high-mass disc galaxies.   

For a more detailed understanding of the effects of different feedback
implementations a direct comparison is required, i.e. different feedback
models applied to the same initial conditions. This was studied, for instance, in the
Aquila comparison project by \cite{2012MNRAS.423.1726S}, who compared
the outcome of simulations with identical initial conditions but 13
models with differing hydrodynamics and feedback schemes. The
differences they found underline the importance of direct code
comparison and might encourage further work in this direction \citep[see also e.g.][]{2005MNRAS.363.1299O, 2010MNRAS.409.1541S, 2011MNRAS.410.2625P}.

In this paper, we present a study of the assembly history of baryonic
matter in two sets of SPH simulations with different feedback
implementations but identical initial conditions. We follow the impact
of the feedback from massive stars right from the beginning of our
simulations until the present day.  Thereby we can study its effect on 
the assembly of the baryons which starts long before they are accreted onto
the galaxy and in many cases may even prevent this. 

The paper is organized as follows: In Section~\ref{simulations} we
describe the simulations and the different feedback models used for
the comparison. Section~\ref{global} contains an analysis of general
differences in the simulations regarding baryon
conversion, angular momentum distribution and mass accretion. In Section~\ref{assembly} we present a more detailed
analysis of the accretion history of gas and stars onto forming
galactic discs, including a study of the angular momentum distribution of the gas. We summarize and discuss our results in Section~\ref{summary}.

\section{Simulations}
\label{simulations}

\begin{figure*}
	\centering
	\includegraphics[width=\textwidth]{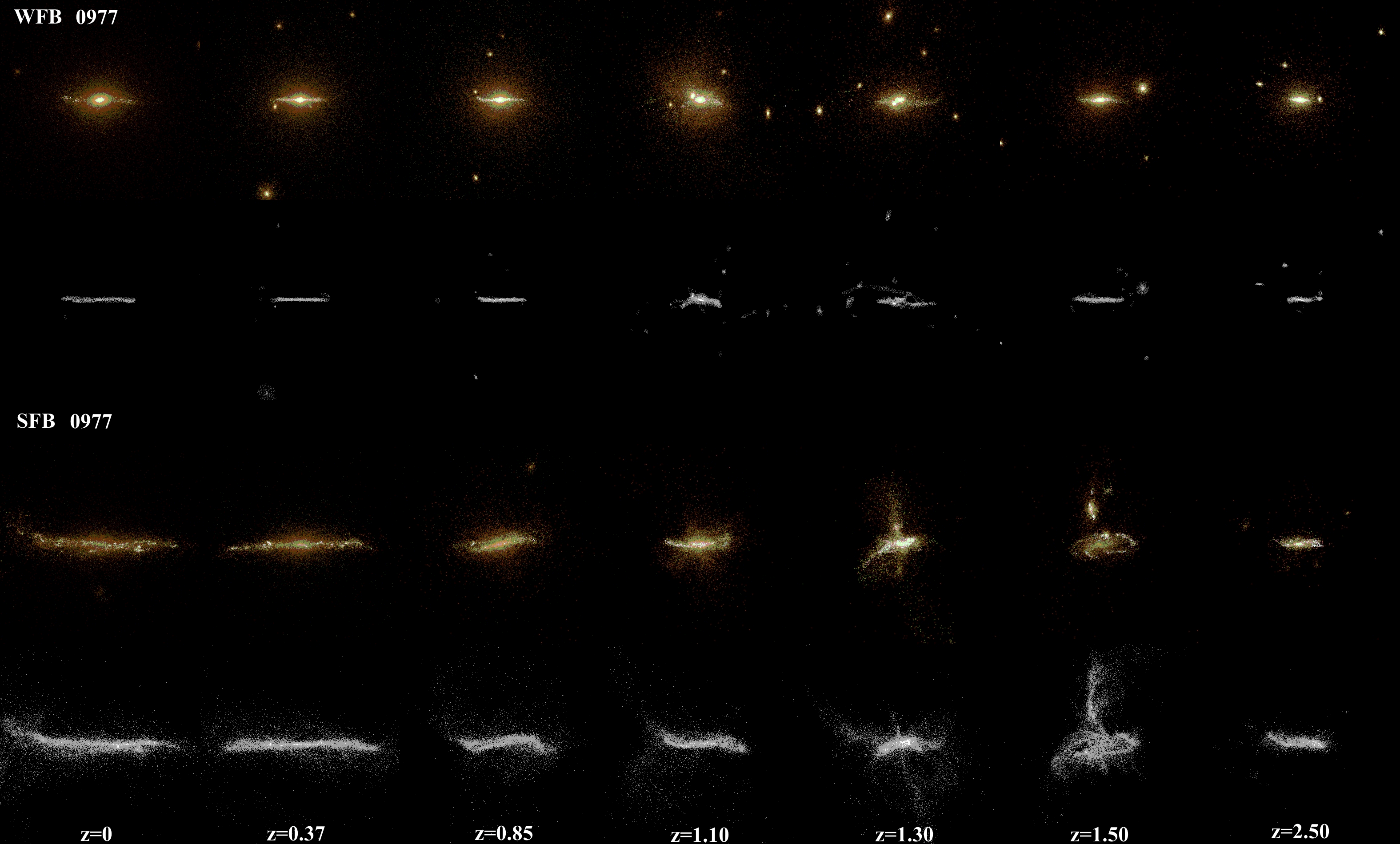}
	\caption{Edge-on mock three-colour $ugr$ images and cold gas densities of the central galaxies in the 0977 models at redshifts $z=0.0; 0.37; 0.85; 1.1; 1.3; 1.5; 2.5$ for WFB (upper rows) and SFB (lower rows). In the SFB model a disc-like structure develops already at early times ($z=2.5$), survives a subsequent merger, and evolves to a flat spiral galaxy by $z=0$. In the WFB model the galaxy undergoes several mergers with smaller satellites which are visible on all images with $z\geq0.37$. At $z=0$, the stellar component of the galaxy has a spheroidal shape, whereas the cold gas forms a thin disc. Each image is 60 kpc wide.}
	\label{morph}
\end{figure*}

We compare two different galaxy formation models, one with weak
feedback favouring the formation of spheroidal galaxies
\citep{2010ApJ...725.2312O} and one with strong feedback supporting
disc formation \citep{2013MNRAS.434.3142A}.

Zoom-in simulations using the weak feedback model were presented in  
\cite{2010ApJ...725.2312O,2012ApJ...744...63O}, where a
description of the numerical details can be found. The model uses a modified
version of the parallel TreeSPH code \textsc{Gadget}-2
\citep{2005MNRAS.364.1105S}, including star formation and supernova
feedback following \citet{2003MNRAS.339..289S} and cooling for a 
primordial gas composed of hydrogen and helium.  Additionally, the
simulations include a redshift-dependent UV background radiation field
with a modified \cite{1996ApJ...461...20H} spectrum.
A subgrid model is used to account for the multiphase structure of the
gas, where clouds of cold gas are embedded in a hot surrounding
medium. Feedback is implemented on a subgrid level as thermal heating
from supernovae, with an energy release of $10^{51}$erg per supernova
that is directly given to the ambient hot gas and in addition
evaporates nearby cold clouds. The heated gas is available for star
formation as soon as it has radiated away enough energy as the
feedback is not explicitly ejective. The masses for the gas and star particles in this model vary from halo to halo over a range $m_{\mathrm{\star,gas}}=2.87 - 7.37 \times
10^{5}\,\mbox{M}_{\odot}$, whereas the dark matter particles have
masses of $m_{\mathrm{dm}} = 1.50 - 3.62 \times 10^{6}\,\mbox{M}_{\odot}$. Each
star particle forms from one gas particle. The comoving
gravitational softening lengths used are
$\epsilon_{\mathrm{\star,gas}} = 200-225\, \mathrm{pc}\,h^{-1}$ for the gas
and star particles and $\epsilon_{\mathrm{dm}} = 445-500\, \mathrm{pc}\,h^{-1}$ for dark matter. From now on, we refer to this model as
`WFB'. 

The second model uses a version of \textsc{Gadget}-3 with a multiphase treatment of gas, and with star formation, metal
production, cooling, turbulent metal diffusion, thermal and kinetic supernova feedback, and
radiation pressure in the neighbourhood of young stars. The basic features of this model were developed by 
\cite{2003MNRAS.345..561M} and \cite{2005MNRAS.364..552S, 2006MNRAS.371.1125S}
and have been extended by \cite{2013MNRAS.434.3142A}, where a detailed
description of the simulations can be found. 
This feedback model assumes
that the released energy per supernova, $10^{51}$erg, is split between kinetic and thermal energy. This energy is
distributed in equal shares to neighbouring hot and cold gas
particles, where cold gas is defined by temperature $T < 8\times
10^{4}$K and density $n > 4\times10^{-5} \mbox{cm}^{-3}$. It should be
noted that the gas particles decouple based on their thermodynamic
properties. The thermal feedback immediately heats hot gas particles,
whereas the energy distributed to cold gas particles is accumulated
until there is enough energy present to turn the cold gas particle
into a hot gas particle. This mechanism produces
large-scale outflows which remove gas from the galaxy so that it is
 not longer available for star formation. Kinetic feedback is
given to the gas particles via momentum transfer. It disperses dense, star-forming regions and can contribute to galactic
fountains. The effect of reducing thermal feedback in favour of kinetic
feedback is weak, whereas the effect of the kinetic feedback on the star
formation efficiency in turn reinforces the thermal feedback. Another
feedback mechanism that is included in this model is radiation
pressure from young massive stars. It is modeled as a continuous
momentum transfer, where the parameters defining the rate of momentum
deposition are chosen in a way that star formation is suppressed in
turbulent environments. These are observed directly in high-redshift
discs \citep[see e.g.][]{2008ApJ...687...59G}. 
Cold gas clumps have been shown to form in some galaxy simulations probably due to unresolved mixing of gas phases in some `classic' versions of SPH \citep[see e.g.][for recent work addressing this issue]{2012MNRAS.427.2224T, 2013MNRAS.429.3353N, 2013MNRAS.434.1849H}. We have used a temperature diagnostic similar to \citeauthor{2012MNRAS.427.2224T} (2012; Fig.\ 13) to test for cold blobs and we have found no evidence for their existence in this model (cf. Fig.~\ref{TdistAqB} in the Appendix). Therefore we assume that our results are not affected by the cold blob issue.
The initial masses for gas and dark matter
particles are the same as in WFB but the masses of gas and stellar particles change during the
simulation through supernova feedback, winds of asymptotic giant
branch stars, and metal diffusion. The comoving gravitational
softening lengths used are $\epsilon_{\mathrm{\star,gas}} = 200-225\,
\mathrm{pc}\, h^{-1}$ for the gas and star particles and
$\epsilon_{\mathrm{dm}} = 450-500\, \mathrm{pc}\, h^{-1}$ for dark
matter. From now on, we refer to this model as `SFB'. 

For the presentation of our results we picked
galaxies that were simulated with these two different feedback models but with the same initial conditions. These were taken from cosmological hydrodynamical simulations of individual
haloes from \cite{2010ApJ...725.2312O} and from the Aquarius Project \citep{2008MNRAS.391.1685S, 2009MNRAS.396..696S}.

The haloes from \cite{2010ApJ...725.2312O} are refined zoom-in simulations of regions from a dark-matter-only
simulation with parameters $h=0.72, \;
\Omega_{\mathrm{B}}=0.044, \; \Omega_{\mathrm{m}}=0.260, \; 
\Omega_{\Lambda}=0.74,$ and $\sigma_8=0.77$. See \cite{2010ApJ...725.2312O} for details.
Our analysis includes the haloes named 0977, 1192, and 1646.
The Aquarius haloes are refined zoom-in simulations of regions from the Millennium II simulation \citep{2009MNRAS.398.1150B} with parameters $h=0.73, \;
\Omega_{\mathrm{B}}=0.040, \; \Omega_{\mathrm{m}}=0.250, \; 
\Omega_{\Lambda}=0.75,$ and $\sigma_8=0.9$.
Our analysis includes the haloes Aquarius B and Aquarius D.
The time resolution for the analysis (time between individual dumps) is $120-140$ Myrs.

In the main part of the paper we present results primarily for halo 0977. Halo 0977 has the second most extended gas and stellar disc of all 19 models of \cite{2013MNRAS.434.3142A} at $z=0$. It was shown in \cite{2014arXiv1401.8164W} that the size and structure of the gas disc are in agreement with observations of gas-rich disc galaxies. 
A concomitant phenomenon of the extended gas disc are the high angular momentum values of the gas. 
0977 is not the cleanest case in terms of quiescent assembly history (like Aquarius B, see appendix) in our sample as it starts undergoing a major merger at $z=0$. However, by choosing this galaxy (and e.g.\ 1646 with misaligned infall) we highlight that many systems with diverse formation histories show the same characteristic assembly features \citep[see also][]{2014arXiv1404.6926A}.
The results for 0977 hold in general for all our haloes.
Selected corresponding figures for haloes 1192, 1646, Aquarius B, and Aquarius D are shown in the Appendix. For noteworthy confirmations or variations of our findings, due to the diverse formation histories of the galaxies, we refer to those haloes in the text.
Characteristic values for all models are summarized in Table~\ref{tab:4}.

We will see in our analysis that the stronger feedback in SFB
significantly affects the gas accretion histories. In this case we find strong
outflows, the gas travels long distances and can leave the galactic
disc for several Gyrs or more. Sometimes the angular momentum of the gas 
changes significantly during this process. In addition, the formation of stars is
strongly suppressed particularly at early times.

\begin{table*}
\begin{tabular}{llrrrrrrrrrrr} 
\hline
\multicolumn{2}{c}{\multirow{2}{*}{model}} &  \multicolumn{1}{c}{M$_{\mathrm{vir}}$} & \multicolumn{1}{c}{R$_{\mathrm{vir}}$} & \multicolumn{1}{c}{M$_{\star}$} & \multicolumn{1}{c}{M$_{\mathrm{gas}}$}  & \multicolumn{1}{c}{\multirow{2}{*}{f$_{\mathrm{gas}}$}} & \multicolumn{1}{c}{\multirow{2}{*}{$\epsilon_{\star}$}} & \multicolumn{1}{c}{\multirow{2}{*}{$z_{1/2}$}} & \multicolumn{1}{c}{$j_{\mathrm{bar}}$} & \multicolumn{1}{c}{\multirow{2}{*}{f$_{\mathrm{eject}}$}} & \multicolumn{1}{c}{$\dot{\mathrm{M}}_{\mathrm{sf}}$} & \multicolumn{1}{c}{\multirow{2}{*}{f$_{\star\mathrm{,rec}}$}}\\ 

&&\multicolumn{1}{c}{\tiny{$\left[10^{11}\mathrm{M}_{\odot}\right]$}} & \tiny{[kpc]} & \multicolumn{1}{c}{\tiny{$\left[10^{10}\mathrm{M}_{\odot}\right]$}} & \multicolumn{1}{c}{\tiny{$\left[10^{10}\mathrm{M}_{\odot}\right]$}} & & & & \tiny{[kpc km s$^{-1}$]} & & \tiny{$\left[\mathrm{M}_{\odot} \mathrm{yr}^{-1}\right]$} & \\ \hline

0977& WFB & $7.1$ & 181 & $6.5$ &  $0.4$ & 0.06 & 0.53 & 2.25 & 472 & 0.07 & 0.50 & 0.16\\  \vspace{1.5mm}
	& SFB  & $12.9$ & 220 & $1.3$ & $1.9$ & 0.60 &0.06& 1.20 & 2171 & 0.62 & 0.93 & 0.55\\
1192& WFB & $12.0$ & 216 & $10.5$ & $0.6$ & 0.03 & 0.51& 2.25 & 318 & 0.05 & 0.01 & 0.04\\ \vspace{1.5mm}
	& SFB & $10.0$ & 202 & $5.0$ & $1.5$ & 0.24 & 0.29 & 0.43 & 1541 & 0.44 & 2.33 & 0.57\\
1646& WFB & $9.5$ & 200 & $7.1$ & $0.3$ & 0.04 & 0.43 & 2.69 & 177 & 0.05 & 0.13 & 0.04\\ \vspace{1.5mm}
	& SFB & $8.1$ & 189 & $1.7$ & $1.4$ & 0.47 & 0.12 & 0.46 & 734 & 0.53 & 0.93 & 0.47\\
AqB & WFB & $6.9$ & 176 & $6.6$ & $0.2$ & 0.03 & 0.57 & 2.69 & 255 & 0.09 & 0.32 & 0.06\\ \vspace{1.5mm}
	& SFB & $7.1$ & 176 & $1.9$ & $1.2$ & 0.39 & 0.16 & 1.08 & 1629 & 0.52 & 0.82 & 0.57\\
AqD& WFB & $17.0$ & 237 & $12.0$ & $0.4$ & 0.03 & 0.42 & 3.32 & 338 & 0.09 & 0.37 & 0.06\\
	& SFB & $14.8$ & 227 & $5.5$ & $0.8$ & 0.13 & 0.22 & 0.65 & 1555 & 0.53 & 1.41 & 0.57\\
\hline
\end{tabular}
\caption{Halo and feedback model, halo virial mass (M$_{\mathrm{vir}}$), virial radius (R$_{\mathrm{vir}}$), galactic stellar mass (M$_{\star}$), galactic gas mass (M$_{\mathrm{gas}}$), galactic gas mass as a fraction of baryonic mass (f$_{\mathrm{gas}}$), baryon conversion efficiency ($\epsilon_{\star}$), half mass formation redshift of the archaeological star formation history ($z_{1/2}$), specific angular momentum of the baryons ($j_{\mathrm{bar}}$), total mass of disc gas which is ejected by $z=0$ as a fraction of the total mass of first accreted disc gas (f$_{\mathrm{eject}}$), $z=0$ star formation rate within the disc ($\dot{\mathrm{M}}_{\mathrm{sf}}$), fraction of $z=0$ disc stars which formed within the disc out of recycled gas (f$_{\star\mathrm{,rec}}$). }
\label{tab:4}
\end{table*}

\section{Global evolution}
\label{global}
 
In this section we investigate the impact of strong stellar feedback
on the morphology and the physical properties of the simulated galaxies and on the assembly history of baryonic matter.
We show a comparison of the morphological evolution of the galaxies and their cold gas densities in halo 0977 in Fig.~\ref{morph} with weak feedback in the upper rows and with strong feedback in the lower rows. The procedure for the creation of these images is described in \cite{2014arXiv1404.6926A}, but here we leave out the dust obscuration for better visibility. Edge-on is defined by the angular momentum vector of the cold gas.
The stellar component of the galaxy in the WFB model has a spheroidal shape that is only slightly flattened at all depicted redshifts. Several small spheroidal satellites are accreted and visible in all images, except at $z=0$. A thin cold gas disc is perceptible from $z=1.5$ to $z=0$.
For SFB both the gas and the stellar components have an extended disc-like shape at all depicted redshifts that is only disturbed for a short time by a merger at $z\sim1.3$. At $z=0$ the galactic disc shows a minor warp that is due to an ongoing merger where the central galaxies are still spatially distinct but they already have some influence on each other.

\subsection{Baryon conversion efficiency}
\label{conv_eff}

\begin{figure}
\begin{center}
    \includegraphics[width=0.47\textwidth]{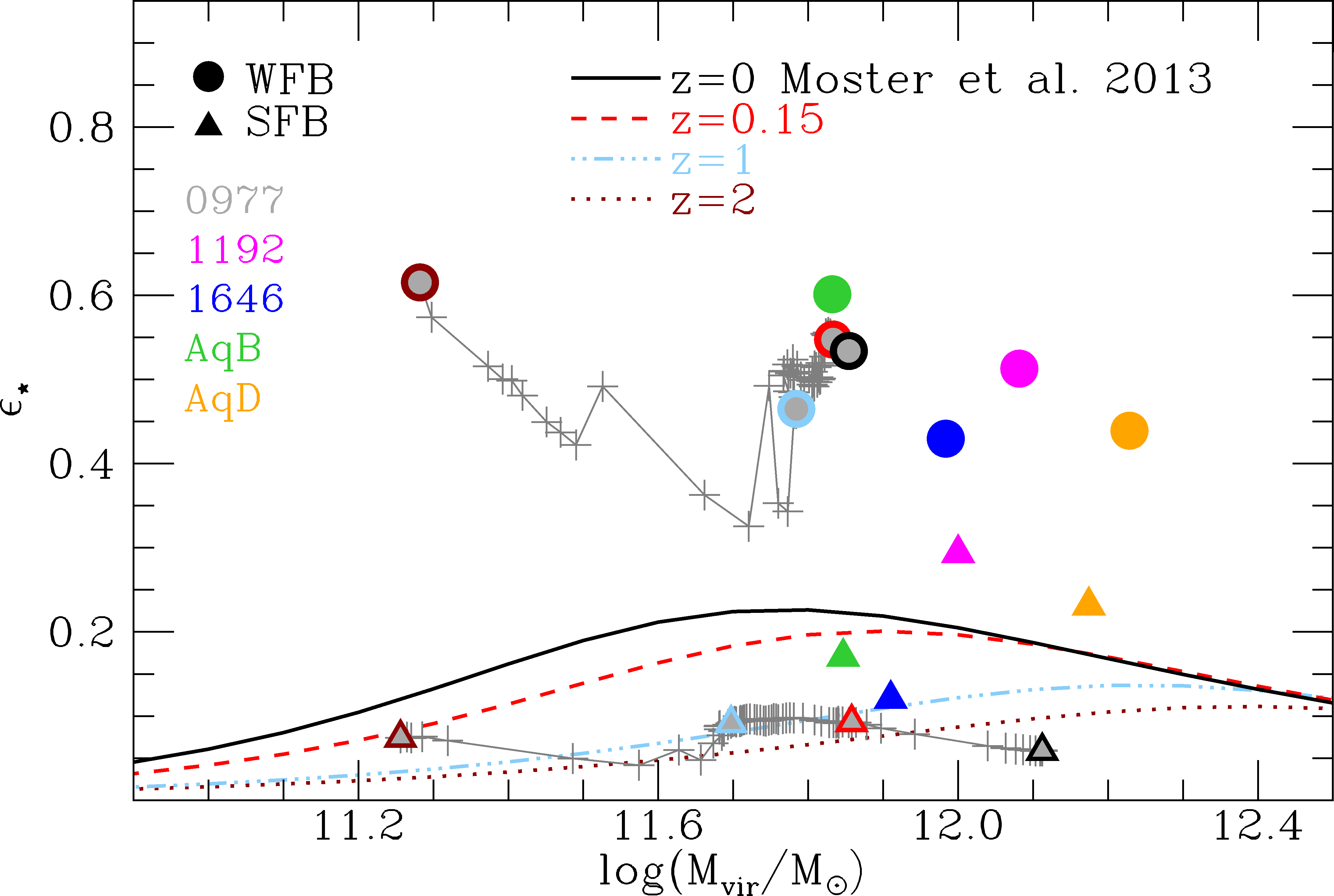} 
  \caption{Baryon conversion efficiency, $\epsilon_{\star}$ as a function of virial halo
    mass for the model with weak (WFB, filled circles) and with strong
    feedback (SFB, filled triangles) at $z=0$. For halo 0977 we add the evolution of $\epsilon_{\star}$ since $z=2$ (grey crosses). Abundance matching estimates at $z=0; 0.15; 1; 2$ (lines) are taken from Moster et al. (2013) and the respective values for halo 0977 are highlighted correspondingly. The SFB model agrees reasonably well with the abundance matching.} 
\label{conveff}
\end{center}
\end{figure} 

The large mass of stars formed already at early times in the WFB model is one of the features distinguishing the two feedback models in Fig.~\ref{morph}. Also, several small spheroidal satellites are accreted and visible in all WFB images for $z>0$, implying that the conversion of gas into stars is more efficient in these objects. 
To quantify this, we take a look at the baryon conversion efficiency, which is defined as $\epsilon_{\star}=\mbox{M}_{\star,\mathrm{central}}/(f_{\mathrm{B}}\cdot
\mbox{M}_{\mathrm{vir}})$, where $\mbox{M}_{\star,\mathrm{central}}$
is the stellar mass of the central galaxy, $\mbox{M}_{\mathrm{vir}}$
is the halo virial mass, and
$f_{\mathrm{B}}=\Omega_{\mathrm{\mathrm{B}}}/\Omega_{\mathrm{m}}$ is
the baryon fraction (see e.g. \citealp{2010MNRAS.404.1111G, 2010ApJ...710..903M}). Stellar
particles are considered to be part of the central galaxy if they lie within r$_{15}$,
i.e. within 15 per cent of the virial radius of the halo,
R$_{\mathrm{vir}}$ (defined as the radius where the mean density
exceeds 200 times the critical density of the universe). 

Looking now at the galactic stellar mass and the associated halo
mass at different redshifts, we find that baryon conversion
efficiencies in the SFB model are significantly lower than for WFB (cf. Table~\ref{tab:4}; \citealp[see also the discussion in][]{2013MNRAS.434.3142A}). This is shown in
Figure~\ref{conveff}, where we plot $\epsilon_{\star}$ as a function of halo mass
for each halo and for both feedback models at $z=0$. For halo 0977 we also plot the evolution of $\epsilon_{\star}$ since redshift $z = 2$ (grey crosses). For comparison,
the expected conversion efficiencies from \cite{2013MNRAS.428.3121M}
are shown for the redshifts $z=0;0.15;1;2$.
  
Regarding halo 0977, the SFB model gives reasonable results for $z=2$ and $z=1$, but at $z=0$ the
conversion efficiency is low. For WFB the conversion efficiencies are
too high at all redshifts. For both models the galaxies experience a major merger 
shortly before $z=1$ (red symbols). 
In the SFB model, there is a
second major merger ongoing at $z=0$, where the haloes have already merged, but the central galaxies are still distinct. Also, the star 
formation rates in SFB are low after $z=1$
\citep[see][]{2013MNRAS.434.3142A}. This explains the low $\epsilon_{\star}$ values in SFB around $z=0$ and the offset in virial
mass between the two models. A companion is also present in the WFB model
but still separated from the halo of the central galaxy. We show
comparative values of $\epsilon_{\star}$ at $z=0.15$, before the last
merger in SFB, when the virial masses of the two galaxies are still
comparable. Even if we add the central stellar mass of the companion
at $z=0$ in SFB, the conversion efficiency is below the expected
value, with $\epsilon_{\star}\approx0.1$ ($\sim 1.25 \sigma$ low). We note that the conversion efficiency in halo 0977 at $z=0$ shows the strongest discrepancy of all models presented in \cite{2013MNRAS.434.3142A} from the predictions of \cite{2013MNRAS.428.3121M}. As apparent from Fig.~\ref{conveff}, all other SFB models discussed in this work scatter more closely around the predictions from abundance matching for $z=0$. Also, the present-day halo masses of the respective WFB and SFB models of each halo are more alike.

\subsection{Star formation history}
\label{sfh}

\begin{figure}
\begin{center}
    \includegraphics[width=0.47\textwidth]{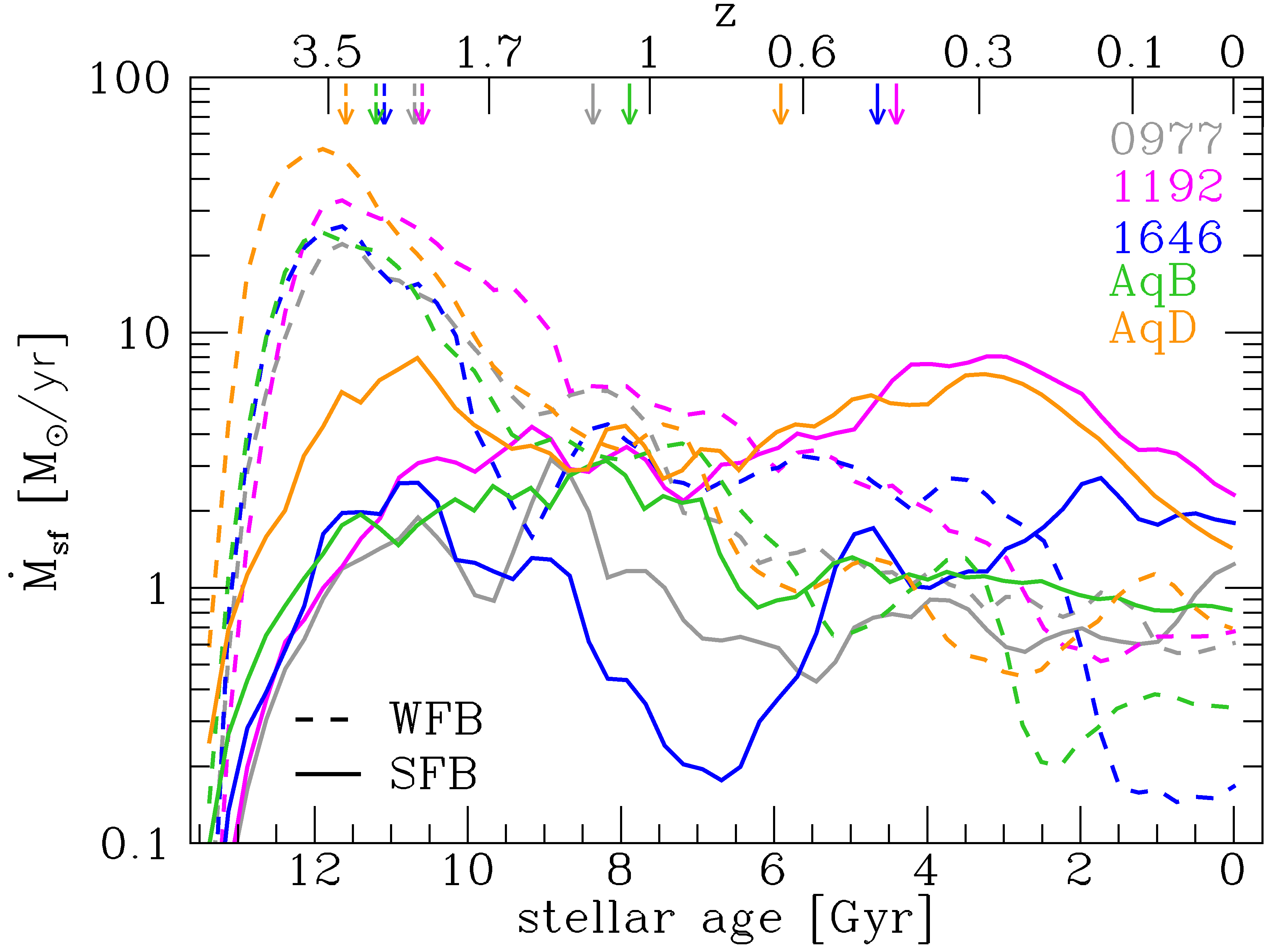} 
  \caption{Archaeological star formation histories for WFB (dashed lines) and SFB (solid lines) for all studied galaxies. The arrows at the upper axis give the half-mass formation redshifts. Half of the stellar mass formed before $z=2$ in WFB, while the curves for SFB are relatively flat with half-mass formation redshifts between $z=1.2$ and $z=0.4$.} 
\label{a-sfh}
\end{center}
\end{figure} 

\begin{figure}
\begin{center}
	\includegraphics[width=0.47\textwidth]{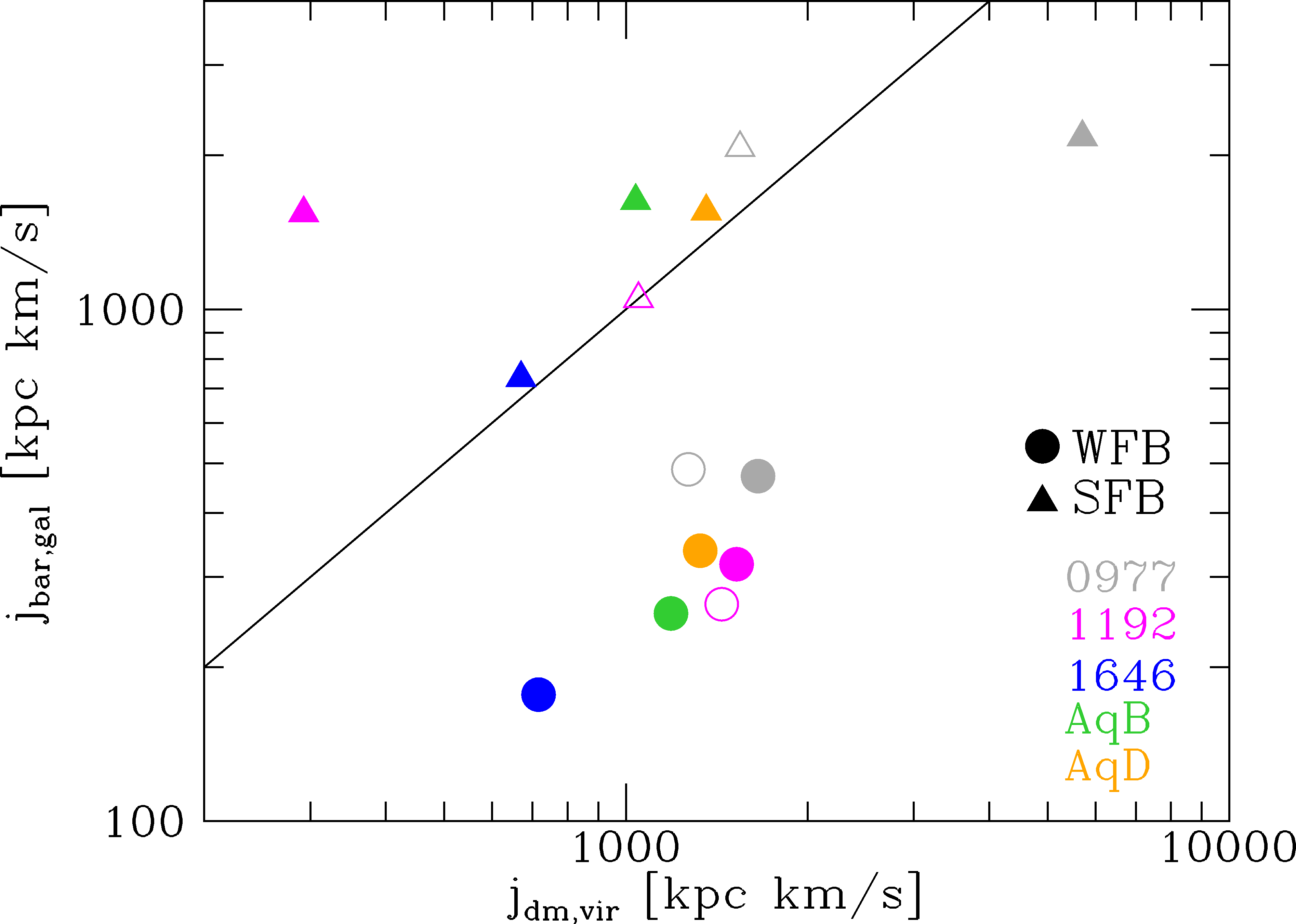}  
  \caption{Specific angular momentum of the baryonic content of the galaxy ($r\le\mathrm{r}_{15}$), $j_{\mathrm{bar,gal}}$ as a function of specific angular momentum of the dark matter halo, $j_{\mathrm{dm,vir}}$ at $z=0$ for WFB (circles) and SFB (triangles) for all studied haloes. The open grey symbols show the $z=0.15$ values for halo 0977; the open magenta symbols show the $z=0.47$ values for halo 1192. The black line gives the 1:1 ratio. The baryons in the SFB model have approximately the virial angular momentum.} 
\label{specific_j}
\end{center}
\end{figure}

As a consequence of the high $\epsilon_{\star}$ values in WFB there are not only more stars, but the stars are also older. This can be seen from the colours of the galaxies in Fig.~\ref{morph}. In the WFB model, old (red) stars dominate the appearance of the galaxy at $z=0$ whereas the SFB model is bluer. The shift in stellar ages towards younger stars in SFB is even stronger for the other haloes, as is shown in Fig.~\ref{a-sfh}. Here, we plot the archaeological star formation histories (i.e. dated by the age of the stars) of all stars within r$_{15}$ at $z=0$ for WFB (dashed lines) and SFB (solid lines) galaxies with the same colour scheme as in Fig.~\ref{conveff}. The half-mass formation redshifts (arrows) are computed using the present-day mass of the stars. Most of the galactic stars in the WFB model are older than 10 Gyrs with half-mass formation redshifts $z\ge2$ for all galaxies (dashed arrows). In contrast, the half-mass formation redshifts for the SFB models shift towards $1.2>z>0.4$ (solid arrows). Although the distribution of stellar ages is not continuous, the curves for SFB are much flatter, which is in agreement with observations of nearby galaxies with masses of $\sim10^{10}\mathrm{M}_{\odot}$ \citep{2004Natur.428..625H}. The fraction of younger stars ($<$ 4 Gyr) is distinctly higher in SFB; in WFB nearly all of the present-day galactic stars formed before $z=1$.
For a more extensive discussion of star formation histories in the SFB model we refer to \cite{2013MNRAS.434.3142A}.

\subsection{Angular momentum distribution}
\label{angmom_dm}

\begin{figure}
\begin{center}
	\includegraphics[width=0.48\textwidth]{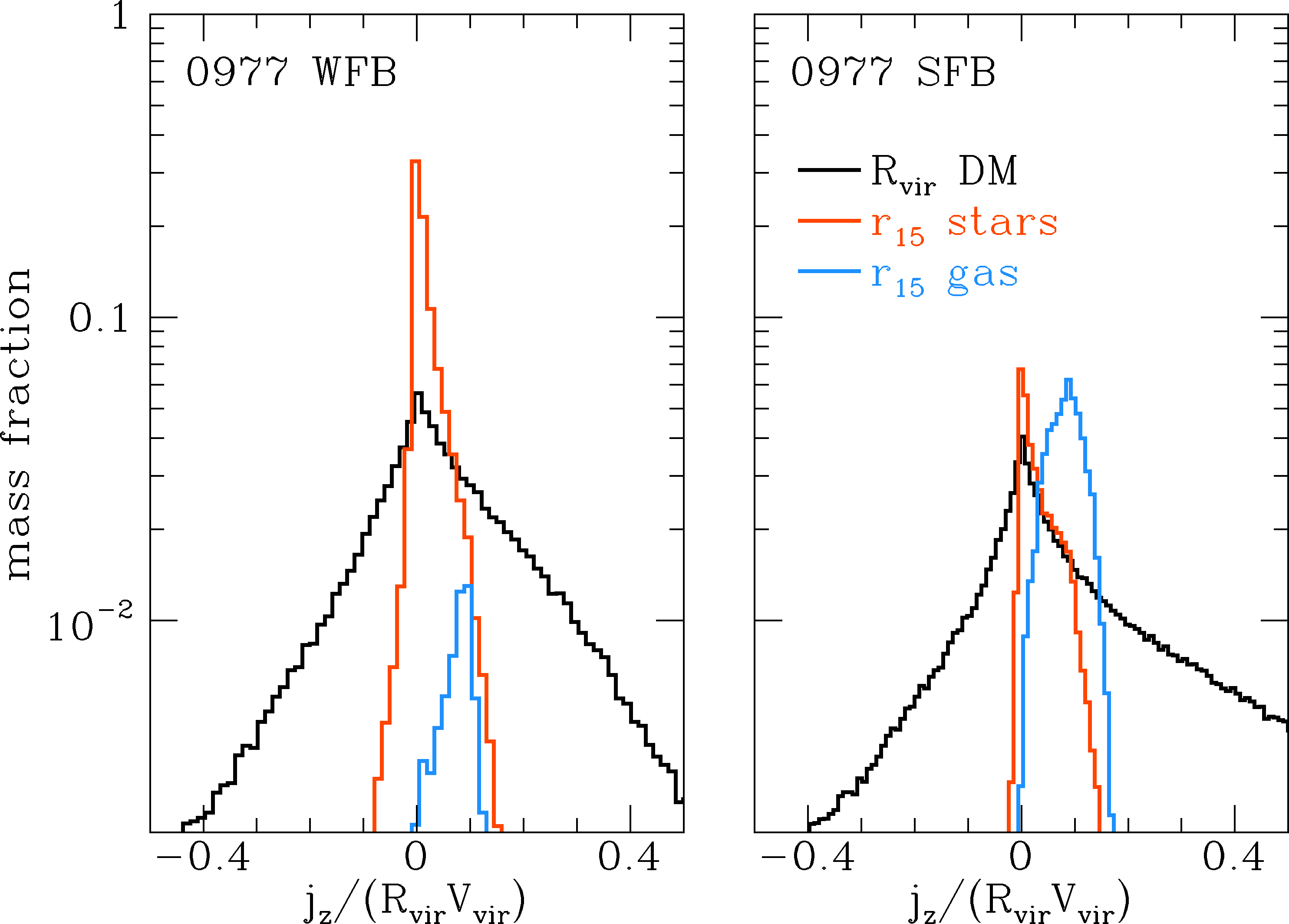}  
  \caption{Distribution of $j_{z}/(\mathrm{R}_{\mathrm{vir}}\cdot\mathrm{V}_{\mathrm{vir}})$ for dark matter within R$_{\mathrm{vir}}$ (black), galactic stars (red) and gas (blue), in the 0977 models for WFB (left panel) and SFB (right panel). The dark matter distribution is normalized to the dark matter mass within the plot range, the lines for galactic gas and stars are normalized to the baryonic mass within r$_{15}$. The stars in SFB are decoupled from the dark matter distribution and fewer stars are counter-rotating.}
\label{halo_acc}
\end{center}
\end{figure}

Besides the stellar content and age distribution of the stars, feedback 
plays an important role in setting the angular momentum content of the simulated galaxies.
Analytical models of disc-galaxy formation suggest that the specific angular momentum of the baryons locked
up in galaxies should be similar to that of their dark matter halos in order to
reproduce the observed scaling relations \citep[e.g.][]{1998MNRAS.295..319M, 2007ApJ...654...27D}. 
We compare in Fig.~\ref{specific_j} the specific angular momentum of the dark matter halo 
(measured within R$_{\rm vir}$), $j_{\mathrm{dm,vir}}$ and that of the baryons for each galaxy (gas as stars
within r$_{15}$), $j_{\mathrm{bar,gal}}$. We use the same colour scheme as
in Fig.~\ref{conveff}, circles and triangles correspond to the WFB and SFB runs, respectively.
For halo 0977 (grey) we also show with open symbols the results at $z=0.15$, just before the merger in the SFB run discussed in Sec.~\ref{conv_eff}, and for halo 1192 (magenta) we show values at $z=0.47$, taking account of halo interaction and misaligned gas infall at later times in the WFB run.

We find a clear segregation between the two models. In the WFB run, galaxies
have only $\sim25$ per cent of the specific angular momentum of the dark matter halo. Instead, the same
objects run with the strong feedback model have comparable or even larger specific angular momenta than
their haloes, as indicated by their location above the 1:1 line.  
This is consistent with the more disc-like morphologies obtained in the SFB runs (see 
Fig.~\ref{morph}). Notice that the strong feedback implementation allows simultaneously
for a small fraction of the baryons to be turned into stars --in agreement with abundance matching
predictions-- but those baryons are able to retain a specific angular momentum similar to that of the halo. This ability of
feedback to efficiently redistribute the angular momentum of the baryons such that the stars
have comparable specific angular momentum to that of the halo is the key to form realistically-looking 
disc galaxies in modern simulations \citep[cf.][]{2011MNRAS.410.1391A, 2013MNRAS.434.3142A, 2014MNRAS.437.1750M}. 

A closer inspection of the angular momentum distribution in these objects reveals
a suppression of counter-rotating stars for the strong
feedback case. That is because the fraction of accreted stars in SFB is low ($\sim15$ per cent ) with most galactic stars formed {\it in situ} out of disc gas particles (see Sec.~\ref{halo_gal_acc} and Table~\ref{tab:3}). Fig.~\ref{halo_acc} shows for halo 0977 
the distribution of $j_z$ at the present day, where $j_z$ is defined as the 
$z$-component of the specific angular momentum after each system has been 
rotated such that the galaxy's total angular momentum is aligned with the $z$-axis. For the alignment of the galaxy, baryonic matter inside three times the half mass radius of the stars in the
galaxy is used. The red and blue histograms show, respectively, the $j_z$ values of the stars and gas within 
the galaxy; black is used for the dark matter halo. The distributions have been normalized
to the baryonic mass within r$_{15}$ for stars and gas, and to the dark matter
mass within the range plotted ($\sim0.9\rm M_{\rm vir}$ for WFB and $\sim0.8\rm M_{\rm vir}$ for SFB). Notice that
roughly $25$ per cent  of the stars in the WFB case are counter-rotating, which is consistent with
the overall low angular momentum content of the system and its spheroid-dominated
morphology in the upper row of Fig.~\ref{morph}. In comparison, this fraction is smaller 
than $15$ per cent  in the SFB run, where most stars co-rotate coherently in a disc (halo 1192 is the most extreme case of our set with $38$ per cent  (WFB) vs.\, $8$ per cent  (SFB) counter-rotating stars). 
The relative heights of the red and blue histograms in Fig.~\ref{halo_acc} reflect the different
contributions of stars and gas to the baryonic content of these objects. Galaxies are
more gas-rich and rotate faster in the strong feedback case. In the following, we investigate 
in detail the origin of the different angular momentum distributions for these two models.

\subsection{Accretion onto the halo and the central galaxy}
\label{halo_gal_acc}

Due to the impact of strong feedback on baryon conversion efficiency, the models vary strongly in the amount of gas and stars that is
accreted onto the haloes. It should be emphasized that `accretion' not only includes accretion of `primordial' gas but also the repeated accretion of cycling gas particles. While this process is of little importance on the halo scale for either model, it becomes a key feature for the accretion onto the central galaxy and the galactic disc in the model with strong feedback.
For halo 0977, about four times more
stars are accreted in the WFB model \citep[see also the weak feedback model of][]{2009ApJ...699L.178N}, but only around $60$ per cent  of the gas mass as compared to
the SFB version.
Still, the fraction of accreted stars is relatively high in the SFB model ($\sim8$ per cent) compared to other haloes in this study because of the ongoing $z=0$ merger. The majority of
the accreted gas in WFB remains in the halo until $z=0$, a smaller
amount is converted into stars, and the rest of the accreted gas,
$18$ per cent , is ejected by $z=0$. In contrast, 36 per cent  
of the accreted gas is ejected in the SFB model by $z=0$. At this point, it should be noted that
particles bound to a substructure that only `flies by' are also
counted as accreted and then ejected.  
The respective mass values are given in Table~\ref{tab:2}.

\begin{table}
\begin{center}
\begin{tabular}{lrrrr} 
\hline
mass $[10^{10} \mbox{M}_{\odot}]$ & WFB & SFB \\ \hline
\textbf{total accreted stars} & \textbf{4.18} & \textbf{0.96}\\
\textbf{total accreted gas} & \textbf{12.13} & \textbf{20.04}\\
(includes repeated accretion events)&&\\
\textbf{first accreted gas} & \textbf{10.35} & \textbf{17.61}\\
\hspace{2mm}\textbullet\hspace{2mm} $\%$ cycling & 11 & 6 \\
\hspace{2mm}\textbullet\hspace{2mm} $\%$ not cycling & 39 & 46 \\
\hspace{2mm}\textbullet\hspace{2mm} $\%$ condensed within R$_{\mathrm{vir}}$ & 32 & 12\\ 
\vspace{1.5mm}
\hspace{2mm}\textbullet\hspace{2mm} $\%$ ejected by $z=0$ & 18 & 36 \\
\textbf{$f_{\mathrm{B}}\cdot$M$_{\mathrm{vir}}$} & \textbf{12.10} & \textbf{21.91} \\
\hline
\end{tabular}
\caption{Respective masses of the different accretion modes onto
  halo 0977. The total accreted gas mass has the expected order of magnitude for the given virial mass at $z=0$ for both models. 
}
\label{tab:2}
\end{center}
\end{table}

\begin{table}
\begin{center}
\begin{tabular}{lrrrr} 
\hline
mass $[10^{10} \mbox{M}_{\odot}]$ & WFB & SFB \\ \hline
\textbf{total accreted stars} & \textbf{4.56} & \textbf{0.29}\\
\textbf{total accreted gas} & \textbf{5.32} & \textbf{14.82}\\
(includes repeated accretion events)&&\\
\textbf{first accreted gas} & \textbf{4.35} & \textbf{7.80}\\
\hspace{2mm}\textbullet\hspace{2mm} $\%$ cycling & 2 & 14 \\
\hspace{2mm}\textbullet\hspace{2mm} $\%$ not cycling & 8 & 9 \\
\hspace{2mm}\textbullet\hspace{2mm} $\%$ condensed within r$_{15}$ & 57 & 22\\
\hspace{2mm}\textbullet\hspace{2mm} $\%$ ejected by $z=0$ & 33 & 55 \\
\hline
\end{tabular}
\caption{Respective masses of the different accretion modes onto the central galaxy (within
  r$_{15}$) for 0977 as shown in Figure~\ref{accmodes}.} 
\label{tab:3}
\end{center}
\end{table}

\begin{figure}
\begin{center}
	\includegraphics[width=0.47\textwidth]{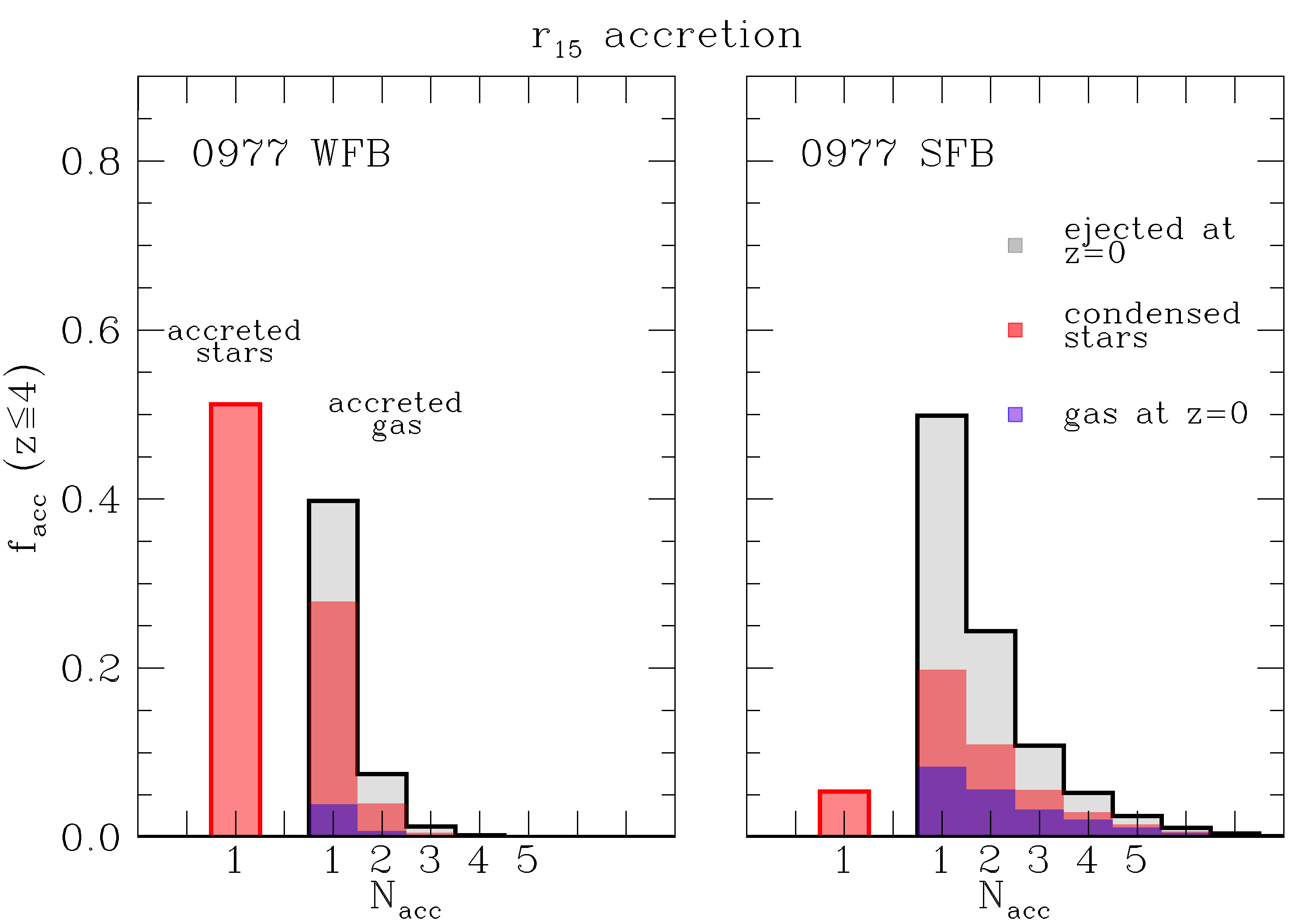} 
  \caption{Fraction of baryons accreted onto the central galaxy (within r$_{15}$) in halo 0977 since $z=4$ as a function of the number of accretion events for WFB (left panel) and for SFB (right panel).
     With stronger feedback direct stellar accretion is less relevant, the mass of gas
     undergoing several cycles grows, star formation is 
    suppressed, and more gas leaves the galaxy by $z = 0$.} 
\label{accmodes}
\end{center}
\end{figure}

The same analysis is performed for baryonic matter that is accreted
onto the central galaxy ($r < \mathrm{r}_{15}$), visualized in
Fig.~\ref{accmodes}, where we plot the star and gas
particles that are accreted onto the galaxy between 
redshifts $z=4$ and $z=0$, f$_{\mathrm{acc}}\, (z\leq4)$, sorted by
how often they have been accreted onto the halo,
N$_{\mathrm{acc}}$. The solid red column on the left
(N$_{\mathrm{acc}}=1$) represents the direct accretion of stars, the
black histogram indicates gas accretion sorted by the number of
accretion events. The blue regions contains all gas particles which are
still a member of the central galaxy by $z=0$, the red region indicates the 
amount of stars that formed from the accreted gas particles within the
central galaxy, and the grey region represents gas particles that were ejected
from the central galaxy by $z=0$. The corresponding masses are given in
Table~\ref{tab:3}. The dominant accretion mode in the WFB model is the
accretion of stellar particles, followed by one-time-only accreted gas 
particles. The direct accretion of stars makes up for about 50 per cent
of all accreted baryonic material since $z = 4$. The fraction of
cycling gas particles which have not been condensed into stars or else have not been ejected by $z=0$ (2 per cent of all accreted gas) is relatively small and most
particles complete only one cycle (N$_{\mathrm{acc}}=2$). The
conversion of gas into stars in the WFB run is very efficient with
57 per cent of the galactic gas particles turned into stars
until $z = 0$. The stars dominate the mass of the central galaxy at
$z=0$. Around 10 per cent of first accreted gas particles are present in the
galaxy at $z=0$, and 33 per cent are
ejected until $z = 0$.  

The behaviour is quite different in the strong feedback model. Direct star
accretion now accounts for only around 5 per cent of the accreted
baryons and the peak of the distribution shifts towards gas
accretion. The dominant accretion mode is single accretion, accounting for $\sim50$ per cent of the total. The fraction of cycling gas
particles which have not been condensed into stars or else have not been ejected by $z=0$ rises to 14 per cent, which accounts for half of the total accreted gas mass (see right panel of
Fig.~\ref{accmodes}). The amount of gas ejected by $z=0$ (grey region) is 55 per cent, but only 22 per cent of the primordial accreted gas particles turn into stars by $z=0$. (The amount of gas that condenses to stars within r$_{15}$ rises significantly for 1192 and Aquarius D, cf. Table~\ref{tab:31192} and Table~\ref{tab:3AqD}.) Much more gas remains in the central galaxy at $z=0$ (Table~\ref{tab:3}) as compared to the week feedback model, both in absolute mass and as a fraction total baryonic mass. This latter fraction is $\sim60$ per cent for SFB but only around $7$ per cent  for WFB. We note that the $z=0$ gas fractions for the SFB models, which range from $13$ to $60$ per cent  (see Table~\ref{tab:4}), are on average too high \citep{2014arXiv1401.8164W}
but are nevertheless closer to observation than the low gas fractions of the WFB models (\citealp[see e.g.][]{2011MNRAS.417.2962P, 2011MNRAS.415...32S} and Tables~\ref{tab:31192}, \ref{tab:31646}, C1, \ref{tab:3AqD} in this paper).

\section{Disc assembly history}
\label{assembly}

For a more detailed investigation of the particular histories of the  
galactic gas particles in the different simulation runs we restrict 
our particle set to `disc particles' by using a measure for the orbital
circularity of individual particles in the galaxy. The circularity of a particle is defined as
$j_{z}/j_{c}$ \citep[c.f.][]{2003ApJ...597...21A,
  2009MNRAS.396..696S}, where $j_{z}$ is the $z$-component of the specific
angular momentum of a single particle and $j_{c}$ is an approximation for the specific angular
momentum of the particle on a circular orbit at the same radius. To
calculate $j_{c}$, a spherically symmetric 
potential is assumed: $j_{c} = r\cdot v_{c}(r)$. Thus, particles with $j_{z}/j_{c} \sim 1$ are on orbits which are close to circular, as expected for
particles associated with a disc. We now pick particles with the
following properties as disc particles: 
\begin{itemize}
	\item $j_{z}/j_{c}=[0.8; 1.2]$\,\, or\,\, $j_{z}/j_{c}=[-1.2; -0.8]$
	\item $|z|<3$\,kpc
	\item $r<0.15\cdot \mbox{R}_{\mathrm{vir}}.$
\end{itemize}
With the possibility of negative values for $j_{z}/j_{c}$, we account
in principle for counter-rotating disc particles, which are, however,
of minor importance in the simulation presented here (but see halo 1646 in Appendix~\ref{1646} that develops a counter-rotating disc). The second
criterion sets a limit to the height above the disc plane up to which
particles are counted as disc particles. The generous choice of $r$
ensures that the particles of the extended discs that form in this
simulation in the SFB model are part of the analysis.
It should be noted that the orientation of the disc and thus the total angular momentum vector can change over several Gyrs of evolution. Due to the very different assembly histories of the galaxies in the two feedback models, these changes evolve in different ways for WFB and SFB. The angle between the total angular momentum of the WFB galaxy and the total angular momentum of the SFB galaxy at $z=0$ is $50^{\circ}$ for halo 0977.

It is clear that the results concerning the galactic disc are in general dependent on disc definition and resolution. However, they differ substantially for the two feedback models due to the details of the feedback implementation.

\subsection{Accretion history}
\label{acc modes}

\begin{figure}
\vspace{0.5cm}
\centering
	\includegraphics[width=0.46\textwidth]{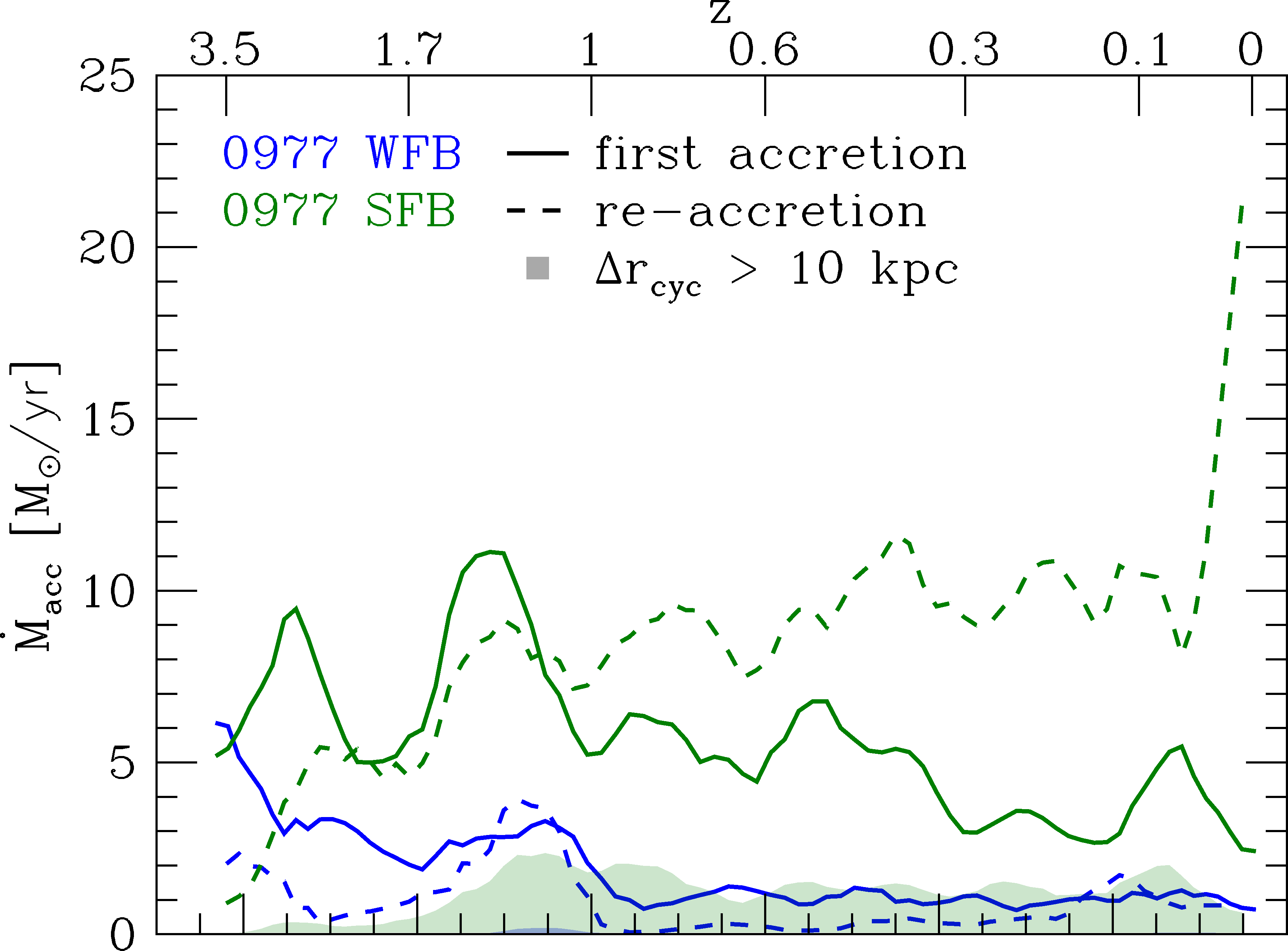} \vspace{1mm}
	\includegraphics[width=0.46\textwidth]{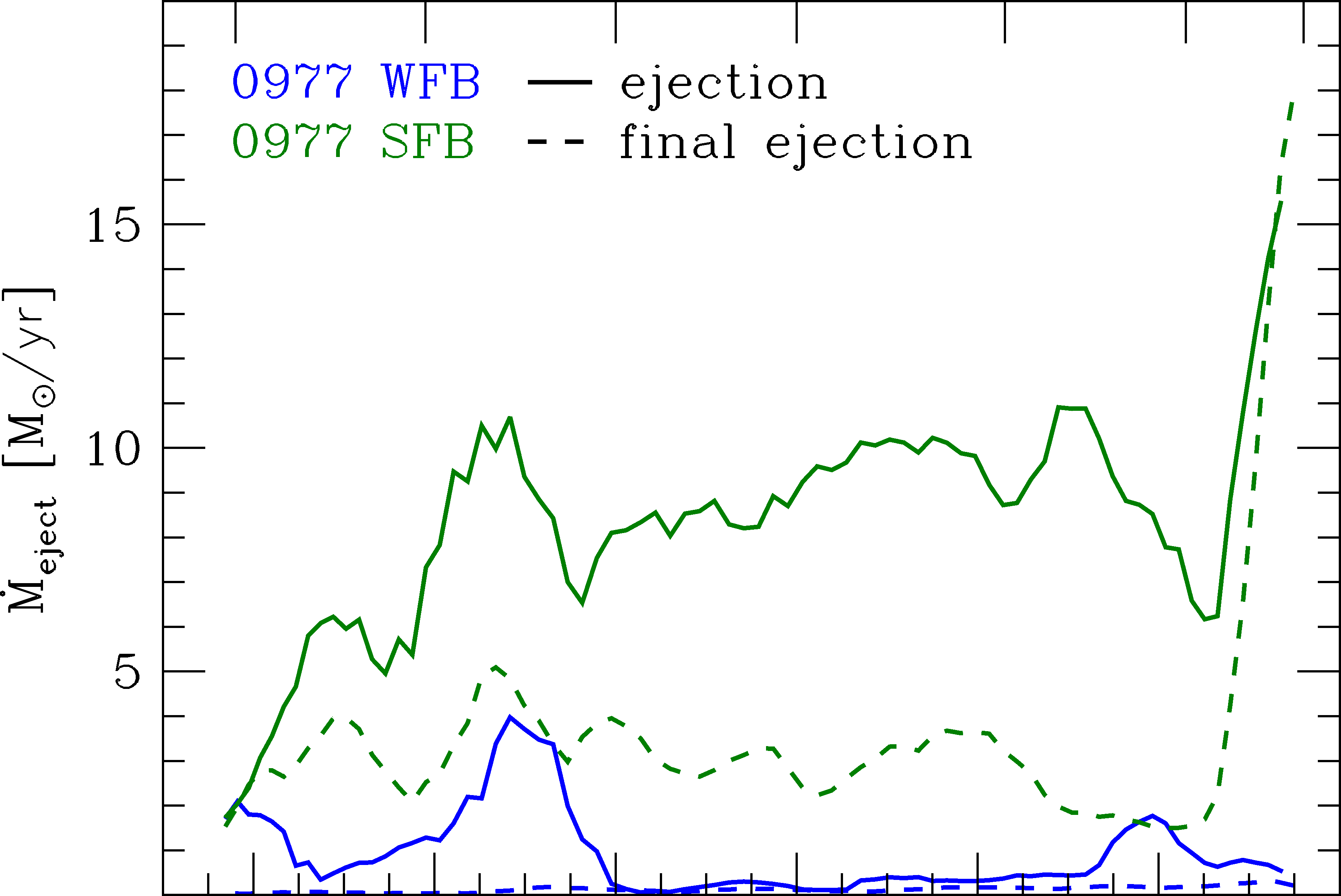}
	\hspace{4mm}\includegraphics[width=0.47\textwidth]{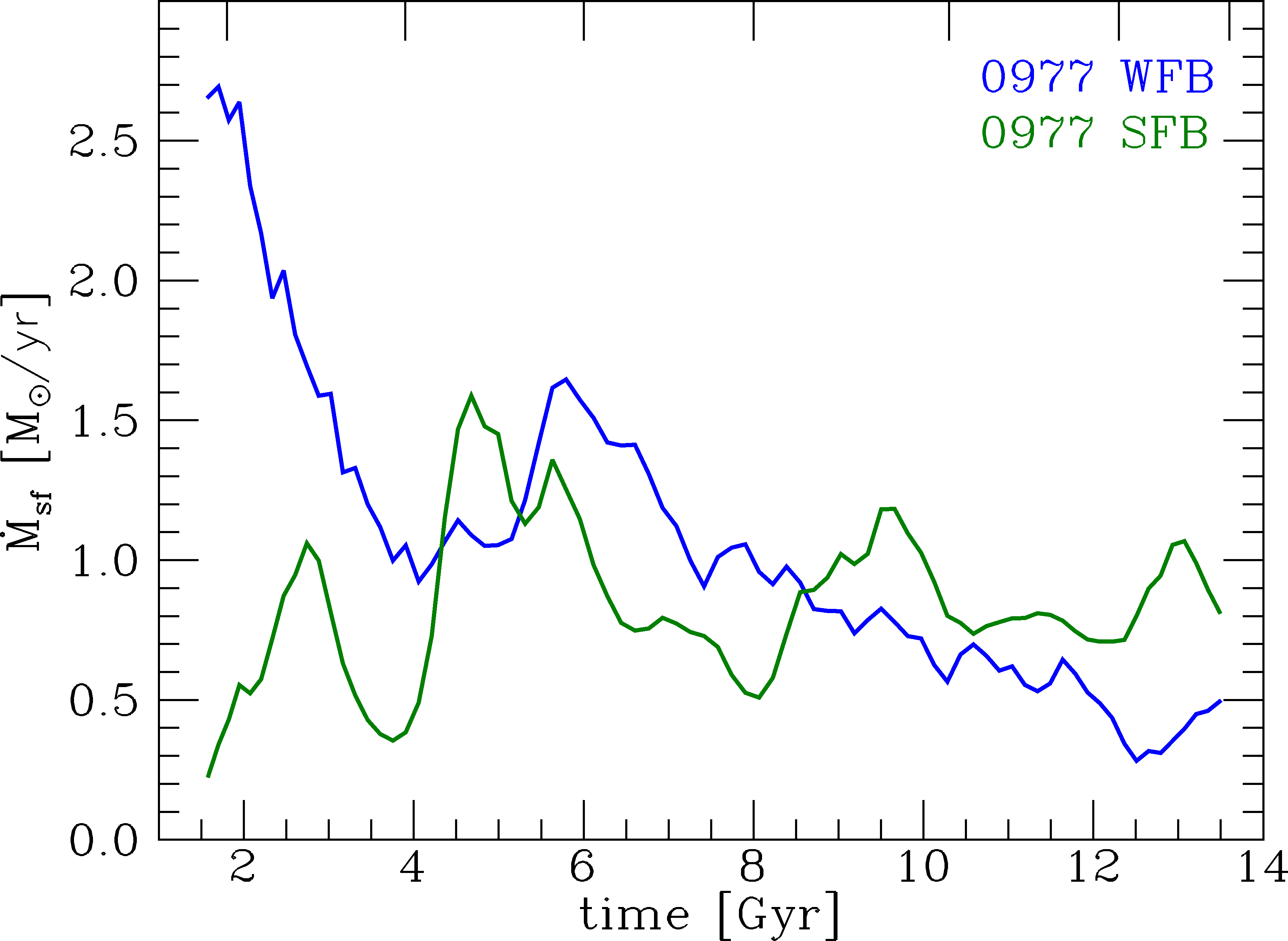} 
  \caption{Top panel: Gas accretion rate onto the disc in the 0977 models separated into first
    accretion (solid lines) and re-accretion (dashed lines; filled
    regions for particles with $\Delta \mathrm{r}_{\mathrm{cyc}} >$ 10
    kpc) as a function of cosmic time and redshift for WFB (blue) and SFB
    (green). Re-accreted particles dominate the accretion rate in SFB
    at late times. 
    Middle panel: Gas ejection rate from the disc separated into ejection
    followed by re-accretion (solid lines) and final ejection (dashed lines) as a
    function of cosmic time and redshift for WFB (blue) and SFB (green). 
    Bottom panel: {\it In situ} star formation rate within the disc as a function of
    cosmic time and redshift for WFB (blue) and SFB (green). Strong feedback
    results in an almost constant star formation rate.} 
  \label{sfr}
\end{figure}

In general, accretion onto the galactic disc is similar to
accretion onto r$_{15}$. Most accreted gas particles form stars in
WFB, and for SFB the number of recycled particles, the average number
of accretion events, and the number of ejected particles by
$z=0$ is significantly larger. Nearly 90 per cent of the particles accreted
onto r$_{15}$ are also accreted onto the galactic disc in SFB, but
only around 40 per cent in WFB, showing again that the disc is less
prominent in the weak feedback model.

In the top panel of Fig.~\ref{sfr} we present the gas accretion histories onto the disc,
$\dot{\mathrm{M}}_{\mathrm{acc}}$ as a function of cosmic time. We separate
gas that is accreted for the first time (solid lines) from gas that is re-accreted (dashed lines). In the
WFB model (blue solid line) the rate of first accretion peaks
at early times at about $6\, \mathrm{M}_{\odot}\mathrm{yr}^{-1}$ and then decreases to a low
level, $\sim1\, \mathrm{M}_{\odot}\mathrm{yr}^{-1}$, after $\sim6$ Gyr. The rate of
re-accreted particles peaks at $5.5\, \mathrm{Gyr}$, then becomes
unimportant, and increases again to a rate similar to the rate of
first accreted particles in the last three Gigayears. 

\begin{figure}
\centering
	\includegraphics[width=0.47\textwidth]{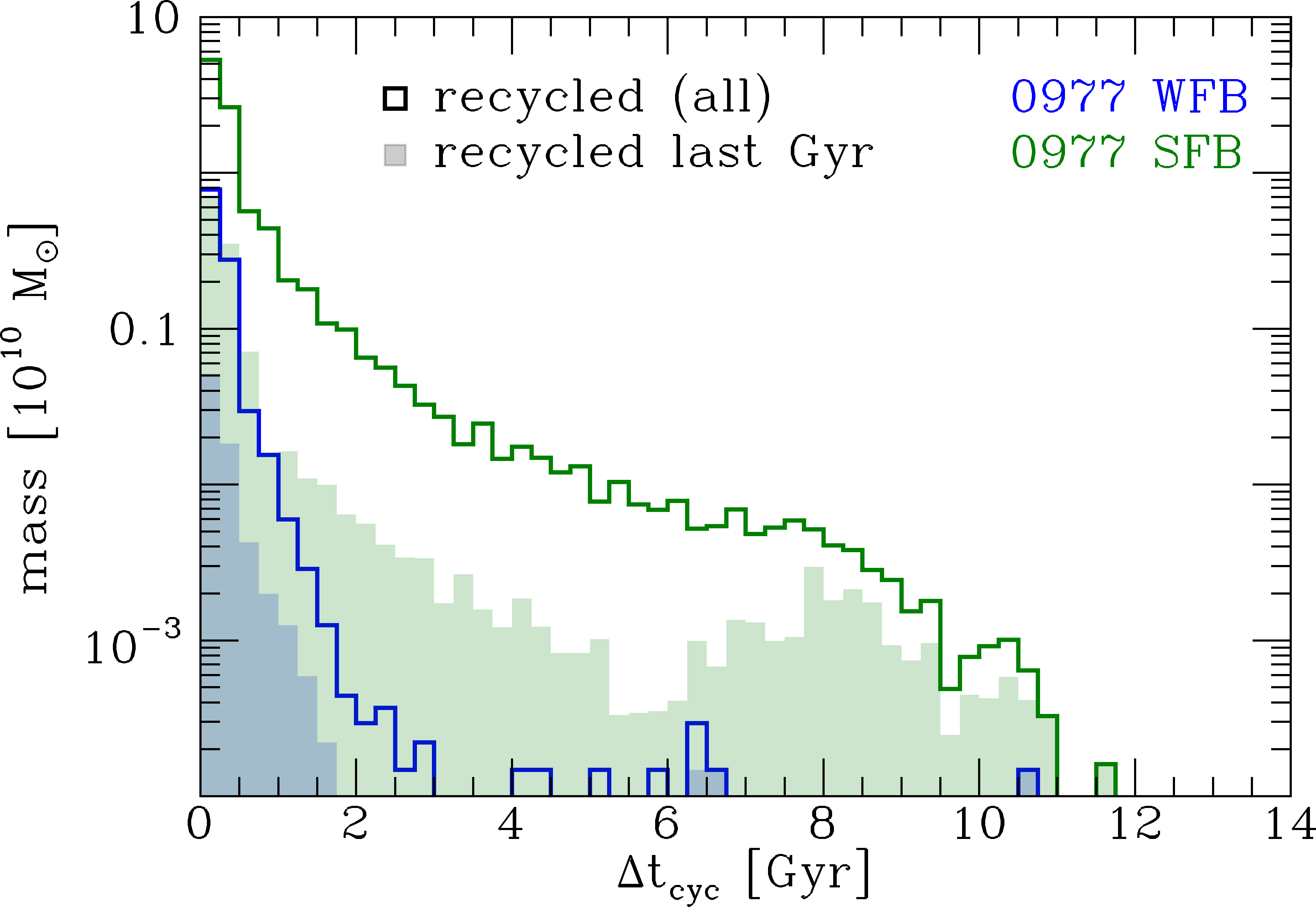}  
  \caption{Recycling times for WFB (blue) and SFB (green) in Gyrs in the 0977 models. Particles which are re-accreted within the last Gyr are shown by the filled areas. There is a strong trend towards higher recycling times in SFB.}
\label{tdelay}
\end{figure}

The situation for the strong feedback model (green) is different and 
the accretion rates are considerably higher. The
peak of first accretion, $\sim12\, \mathrm{M}_{\odot}\mathrm{yr}^{-1}$, is shifted 
towards $\sim5\, \mathrm{Gyr}$. At this time the galaxy experiences
a major merger (cf. Fig.~\ref{morph}, images at $z=1.5; 1.3$) which is barely visible for the WFB
model. This merger influences the later accretion history
of the SFB galaxy (see Fig.~\ref{tdelay}). After the merger, accretion
stays at a relatively high level ($\sim5\, \mathrm{M}_{\odot}\mathrm{yr}^{-1}$) after
$\sim6\, \mathrm{Gyr}$. In the 
SFB model first accretion is primarily in the form of gas rather
than stars (see Fig.~\ref{accmodes}). Even more interesting is the fact the re-accretion of gas starts to dominate after $\sim6\,
\mathrm{Gyr}$, underlining the importance of gas recycling for 
the evolution of the galaxy in this model. 
The re-accretion peak at $z=0$ is partially caused by the tidal elongation and the warping of the galactic disc through the ongoing merger (cf. Fig.~\ref{morph}), but we note that the re-accretion of gas at late times is generally dominating in the SFB models (cf. e.g. Fig.~\ref{gasrateAqD}).
The re-accretion rates of
particles with travel distances greater than 10 kpc are indicated by
the filled regions
(cf. Sec.~\ref{trav} for a definition and discussion), indicating
that most of the gas cycles close to the galaxy in a galactic
fountain. 

The re-accretion rates closely follow the ejection rates, which are shown
in the middle panel of Fig.~\ref{sfr} as solid lines. Here, particles that have been ejected and
did not return by $z=0$ are indicated by the dashed lines. Whereas
almost no gas is ejected in WFB, there is nearly constant mass loss for the SFB model at $\sim3 \mathrm{M}_{\odot}\mathrm{yr}^{-1}$ (again, the peak at $z=0$ is
related to the ongoing merger in the SFB case). 
We quantify the difference in gas ejection vs. accretion in the two models by measuring the total mass of disc gas particles which are ejected by $z=0$ as a fraction of the total mass of first accreted disc gas particles, f$_{\mathrm{eject}}$. With f$_{\mathrm{eject}}=0.62$ in the SFB model, this value is $\sim9$ times higher than for the WFB model, underlining the effect of the strong feedback model on gas ejection (see Table~\ref{tab:4} for the respective values for the other haloes).

In the bottom panel of Fig.~\ref{sfr} we plot
the {\it in situ} disc star formation rate, $\dot{\mathrm{M}}_{\mathrm{sf}}$
as a function of cosmic time. We find higher formation rates at earlier
times for WFB, reflecting the high baryon conversion efficiencies in
the weak feedback model, and higher formation rates at later time for
SFB that are fed by the (re-)accreted gas particles (see Table~\ref{tab:4}). 
If we compare
the ratio of the gas ejection rate to the {\it in situ} star formation rate,
the mass loading factor $\eta$, we get significantly higher values in SFB ($\eta \ge 10$) than in WFB ($\eta \sim 1$)
at all times, consistent with the lower conversion
efficiency. In the strong feedback model, the mass of 
ejected material is at all times higher than the mass formed in stars,
but the gas reservoir in the disc is constantly re-filled by recycled
gas. 
To quantify the correlation between {\it in situ} formed stars and recycled disc gas \citep[cf. also][]{2013arXiv1306.5766B}, we measure the percentage of $z=0$ stellar disc particles which have formed within the disc from recycled gas, f$_{\star\mathrm{,rec}}$. Whereas we find f$_{\star\mathrm{,rec}}=0.55$ for the strong feedback model, the corresponding value in the WFB model is only 0.16. We note that this is also the highest fraction in the sample for the WFB model (cf. Table~\ref{tab:4}). This should have significant consequences for the chemical evolution of the disc.

\begin{figure}
\centering
	\includegraphics[width=0.47\textwidth]{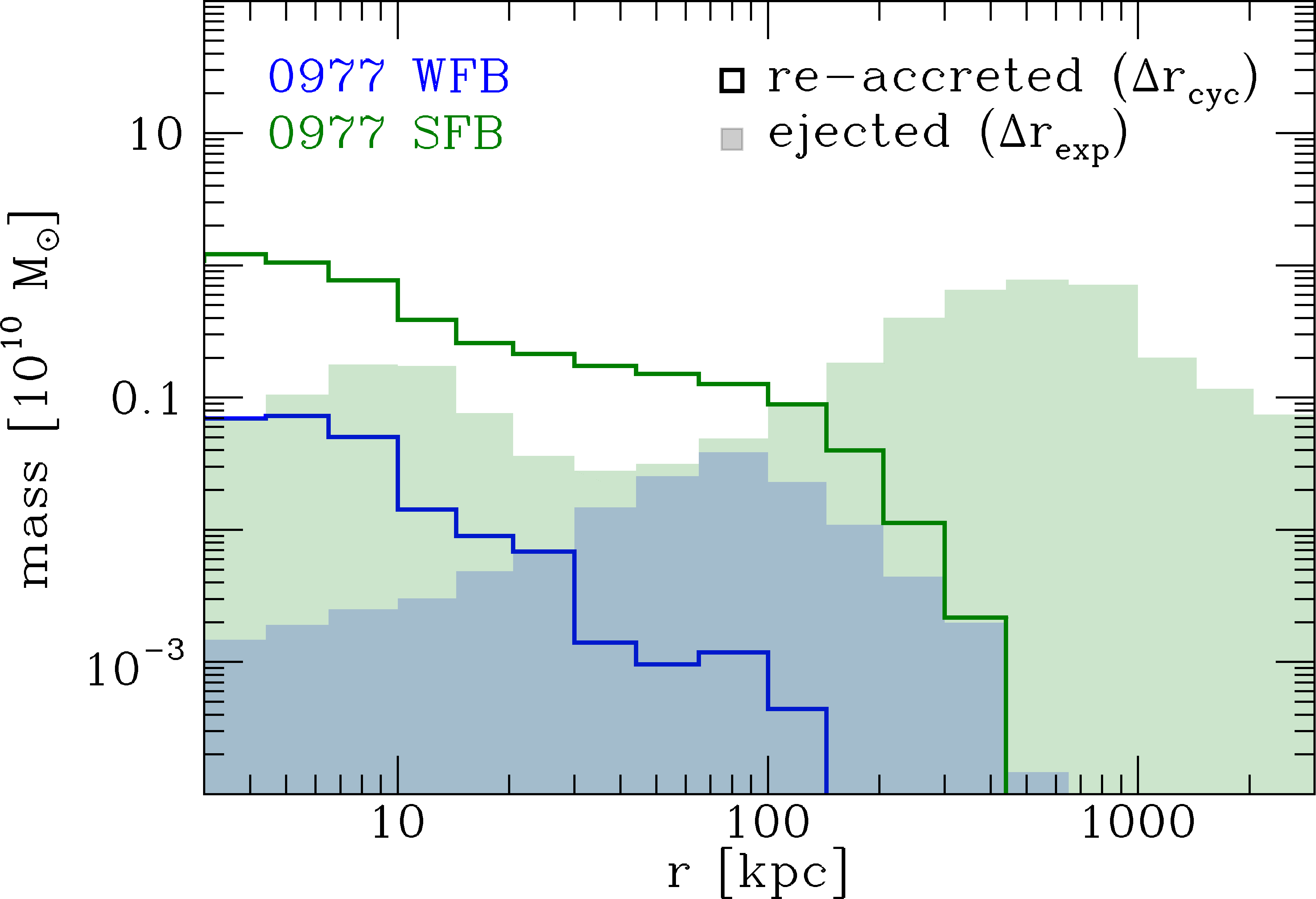} 
  \caption{Travel distances for WFB (blue) and SFB (green) for cycling particles (solid line histograms) and for particles which are ejected by $z=0$ (filled histograms) in the 0977 models. The typical travel distance for cycling particles is higher in SFB, and by $z=0$ expelled particles reach higher distances in SFB.}
\label{mdist}
\end{figure}

\begin{figure*}
\centering
	\includegraphics[width=0.78\textwidth]{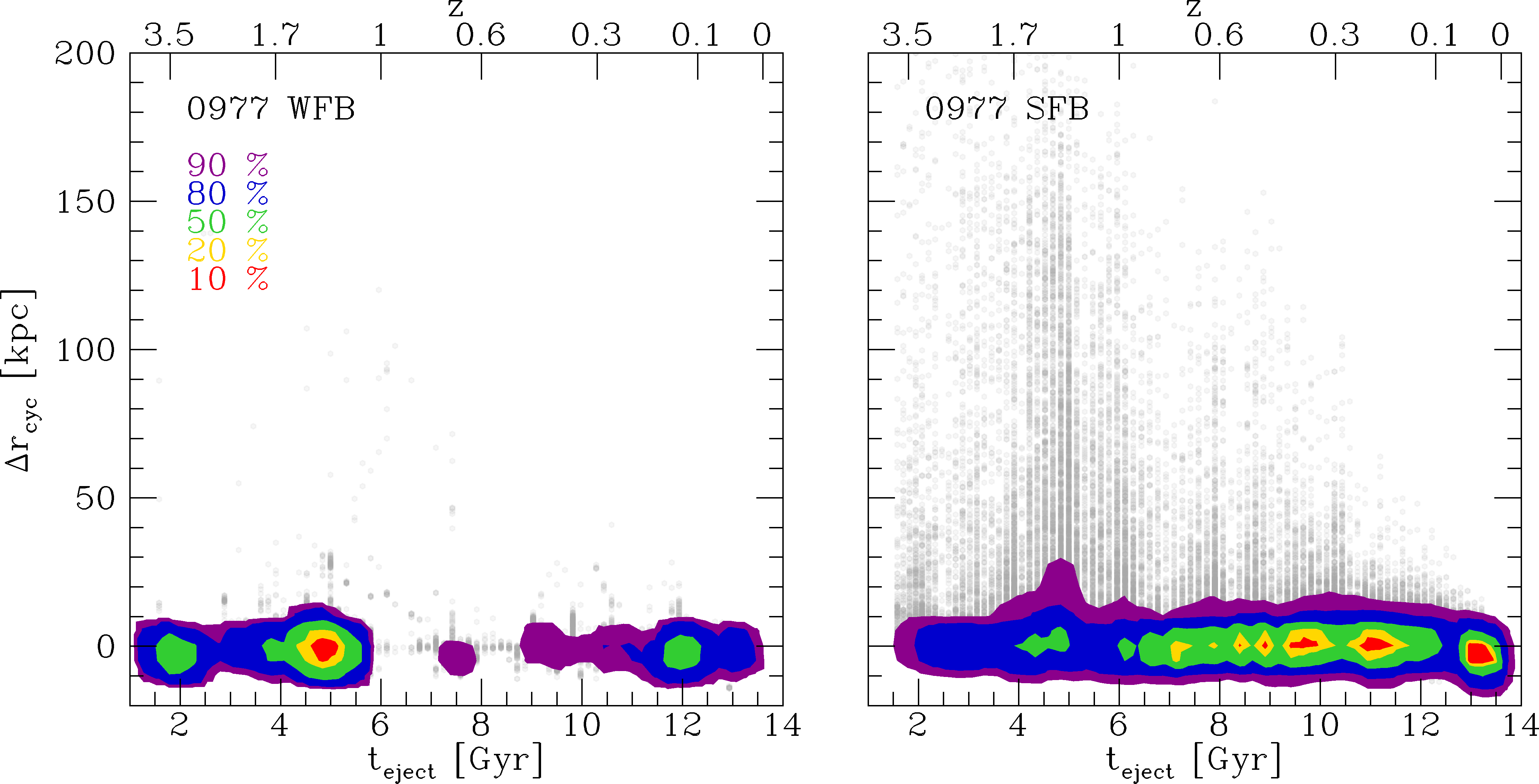} 
  \caption{Travel distance
    as a function of ejection time for WFB (left panel) and SFB (right
    panel) in the 0977 models. Individual particles are shown as grey dots,
          the coloured regions indicate the respective fractions at the
          indicated levels. In SFB, the particles travel farther and merger events are reflected by travel distance peaks.} 
\label{mdist_time}
\end{figure*}

\subsection{Recycling times and travel distances}
\label{trav}
In Fig. \ref{tdelay} we plot the recycling time (or delay time), $\Delta$t$_{\mathrm{cyc}}$, indicating how long a 
particle is out of the disc before being re-accreted again. Using a
different feedback model it has been argued that most of the gas
is re-accreted after 1-2 Gyrs
\citep[cf. also][]{2010MNRAS.406.2325O, 2013arXiv1306.5766B}. This is also true for our models (solid lines in Fig.~\ref{tdelay}). For both, WFB and SFB, the
recycling times peak at $\Delta$t$_{\mathrm{cyc}} < 1$ Gyr but the SFB
model shows a prominent extended wing towards longer delay times, up to 11 Gyrs. The filled histograms indicate the delay time
distribution of gas particles which have been re-accreted within the
last Gigayear of galactic evolution. It is interesting to see that some gas
expelled from the galaxy, in a merger event more than 8 Gyrs ago, has
been accreted only recently. This association of merger activity and recent re-accretion of gas can also be seen nicely in Fig.~\ref{tdelay1646}, where we show the same plot for halo 1646. We find recent re-accretion peaks for gas with $\Delta$t$_{\mathrm{cyc}}\sim2$ and $\sim4.5$ Gyrs that was ejected in merger events.

We define the maximum travel distance of the cycling gas particles as the difference of the maximum distance of the particle from
the galactic center during the cycle and the distance it had from the galactic center right
before its ejection,
$\Delta$r$_{\mathrm{cyc}}=\mathrm{r}_{\mathrm{max}} -
\mathrm{r}_{\mathrm{eject}}$. Therefore, cycling particles can have  
negative values of $\Delta$r$_{\mathrm{cyc}}$ if they migrate inwards
and are not moved away from the galactic center more than
r$_{\mathrm{eject}}$ during the cycling. This applies to nearly half
of the cycling particles in WFB and to around 20 per cent of the
particles in SFB. In the strong feedback model about an order of
magnitude more gas (at all radii) is in the recycling mode. Nevertheless, a very high
fraction of the cycling gas in both models stays within 10 kpc, namely 97 per cent for WFB and 86 per cent for SFB, as can be seen in the top panel of 
Fig.~\ref{sfr} as well as in Fig.~\ref{mdist}. In the 
SFB model, however, some of the cycling particles can
travel as far as 300 kpc, i.e. beyond the virial radius, whereas
in the weak feedback case this mode is limited to the inner 150 kpc. 
Additionally, we define the maximum travel distance of particles which are expelled from the galactic disc by $z=0$ as the difference of the maximum distance of the particle from
the galactic center after its ejection from the galactic disc and the distance it had from the galactic center right before its ejection, $\Delta$r$_{\mathrm{exp}}=\mathrm{r}_{\mathrm{max}} -
\mathrm{r}_{\mathrm{eject}}$.
The escaping gas (filled regions in Fig.~\ref{mdist}) in SFB can travel
much larger distances indicating that with strong feedback a single
galaxy can have contributed to metal enrichment in regions that are more than 1 Mpc away.
   
In addition, for SFB the gas travels significantly larger distances at
earlier times, as compared to the WFB model. Certainly, larger travel distances at earlier times are in part caused by the fact that recently ejected particles cannot have reached large distances before being re-accreted due to limited velocities (but see Fig.~\ref{mdist_time1646} for an example of higher travel distances at later times, due to strong merger activity). In Fig.~\ref{mdist_time} the travel
distances of the cycling gas particles are plotted as a 
function of cosmic time. In the SFB model (right panel) there is a
concentration of cycling particles with relatively high travel
distances between 4 and 5.5 Gyr. This peak is associated with the
aforementioned  merger event and some of the ejected gas particles
need more than 8 Gyrs to return to the galactic disc. They can be
seen in Fig.~\ref{tdelay} as a part of the bump around 8 Gyr made up of
those particles in SFB that are accreted within the last Gigayear. Those signatures of merger events can be seen more distinctly in some of the other haloes, cf. Figs.~\ref{mdist_time1192}, \ref{mdist_time1646}.

\begin{figure*}
\centering
\vspace{1cm}
	\includegraphics[width=0.78\textwidth]{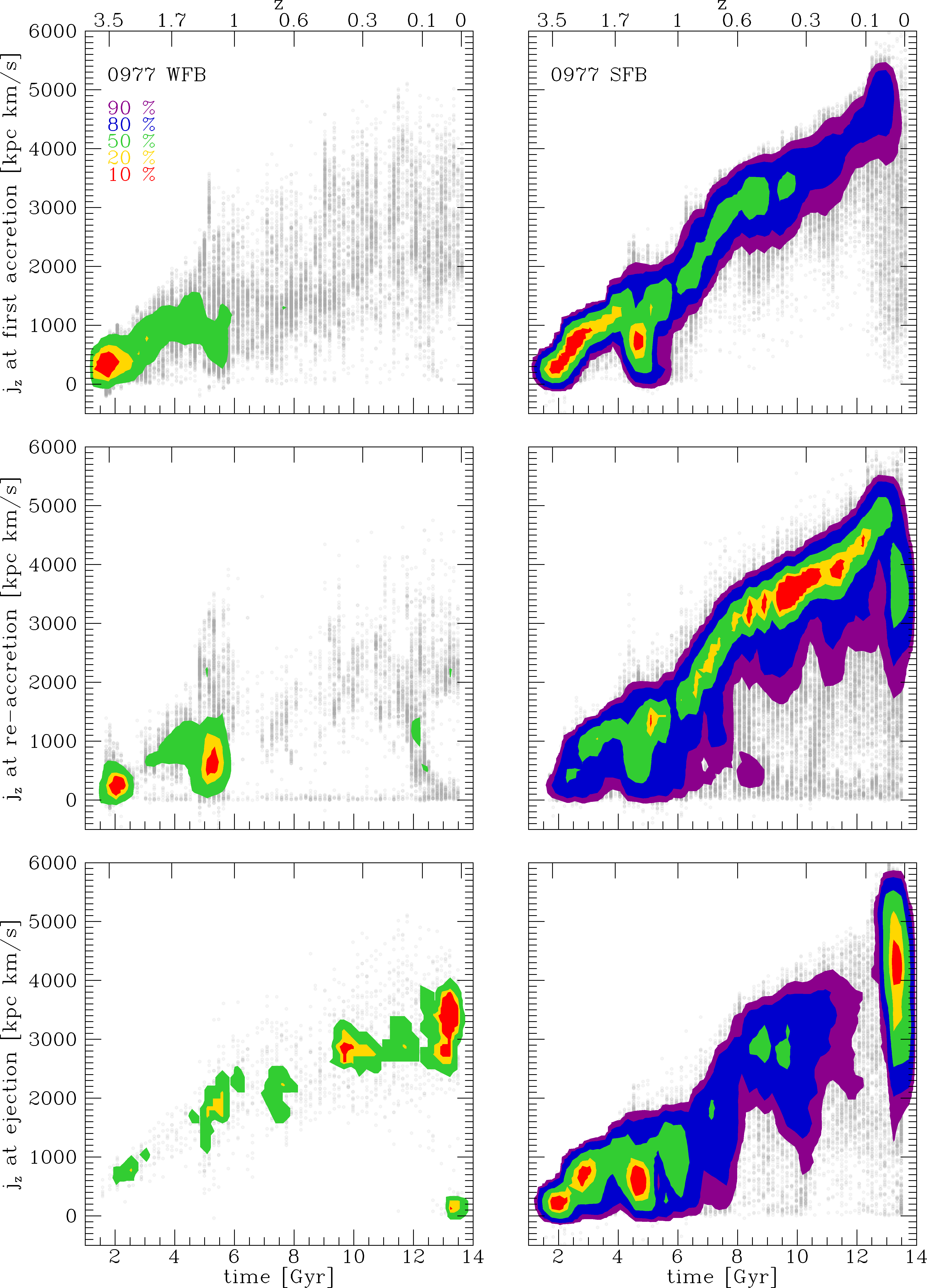}
  \caption{Top panel: Angular
          momentum in the $z$-direction for first-time accreted particles (onto the disc) as a
          function of accretion time for WFB (left panel) and SFB
          (right panel) in the 0977 models. There is a clear trend for more late
          accretion of higher angular momentum gas in SFB as compared to WFB. This promotes disc
          formation.
          Middle panel: Same as the top panel but for re-accreted particles as
    a function of accretion time for WFB (left panel) and SFB (right
    panel). In contrast to WFB, accretion of recycled particles is the dominant accretion mode after $z\sim 1$ for SFB. This is mainly driven by star formation activity, as most particles cycle close to the galaxy. This accretion mode also promotes the formation of galactic
    discs.
          Bottom panel: Same as the top panel but now for particles that are
    ejected from the disc by $z=0$ as a function of time of ejection for WFB (left panel) and
    SFB (right panel). With strong feedback (right, SFB), a
    significant fraction of the gas is ejected at early times, with
    low angular momentum. Again, the early ejection of low angular
    momentum promotes disc formation.}  
\label{jej}
\end{figure*}

\begin{figure*}
\centering
	\includegraphics[width=0.78\textwidth]{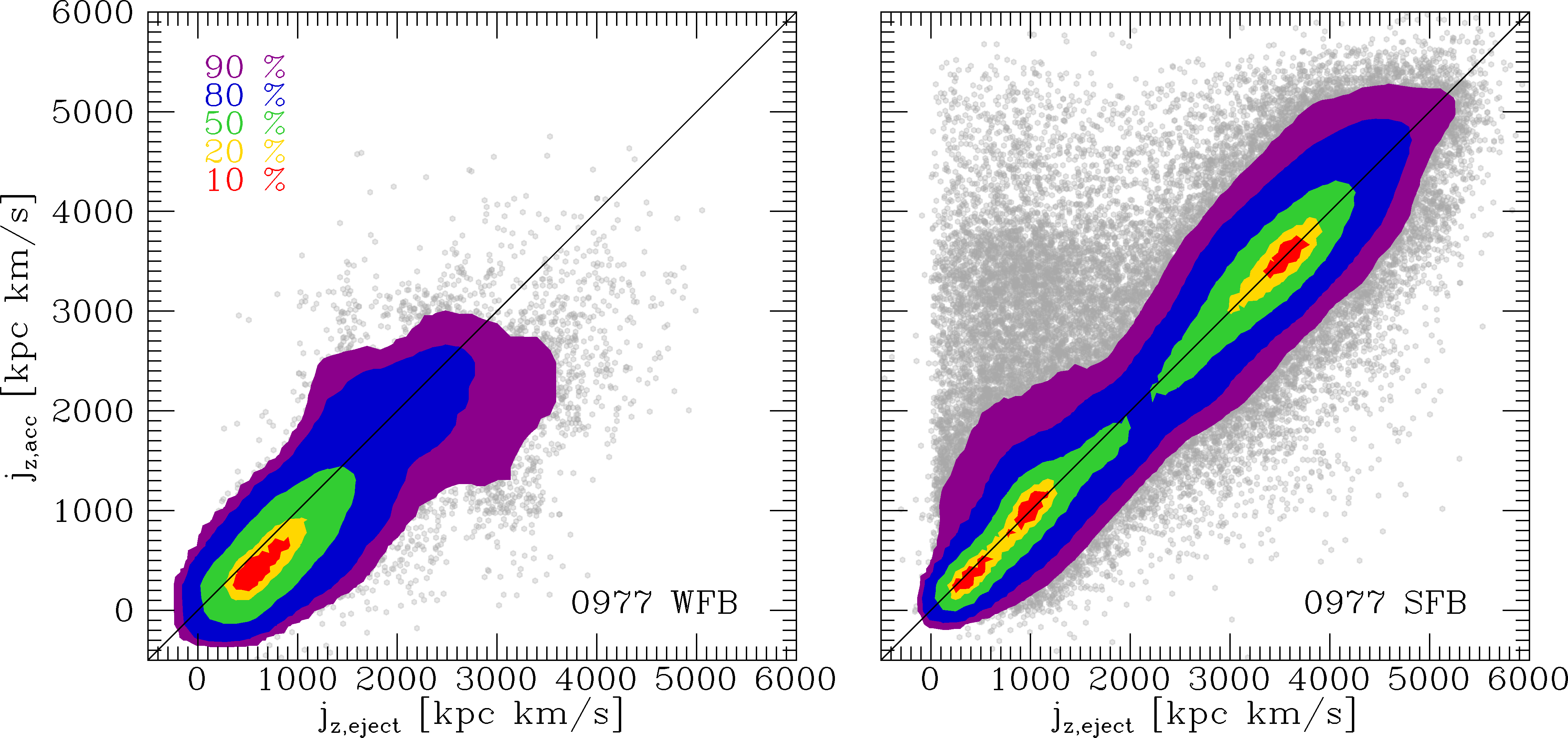} 
   \caption{Angular momentum at ejection, $j_{\mathrm{z,eject}}$
     vs. angular momentum at accretion, $j_{\mathrm{z,acc}}$ for 
     cycling gas particles for WFB (left panel) and for SFB (right
     panel) in the 0977 models. In general, the cycling gas in SFB has higher angular
     momentum both at ejection and accretion. Most particles keep
     their angular momentum. Apparently, some fraction of the gas
     particles in SFB (right) can gain angular momentum (as opposed to
     some loss for WFB), in particular when ejected at low angular
     momentum.}  
\label{joutrec}
\end{figure*}

\subsection{Evolution of angular momentum}
When a gas particle enters the galactic disc its angular momentum $j_{z}$ is
recorded. For the SFB model (right panel in the top row of Fig.~\ref{jej}) we find
that the angular momentum of first accreted gas particles continuously
increases with time, consistent with a cosmological inside-out growth
(e.g. \citealp{1998MNRAS.295..319M}). This might be delayed by merger activity (cf. e.g. top panel of Fig.~\ref{jej1192}). The concentration of first
accreted particles with relatively low 
angular momentum ($<$ 1000 kpc km s$^{-1}$) around 5 Gyr is again a 
clear signature of the merger event, which is also visible for WFB in
the left panel. The steady growth of angular momentum of first
accreted particles with time is to some extent also detectable in WFB,
but it is much weaker and leads to the old problem of the
angular momentum catastrophe \citep{1991ApJ...380..320N}. For both
models, first accretion peaks at early times (the first 6 Gyrs, see
also the top panel of Fig.~\ref{sfr}). The middle panels of Fig.~\ref{jej} show the same analysis for
re-accreted particles. In WFB there is little late re-accretion,
whereas re-accretion of high angular momentum gas dominates the
accretion for SFB. Note that this phenomenon is connected to the fact that for halo 0977 at late times in the strong feedback model stars are forming almost exclusively at large radii (r $>$ 10 kpc) and there is almost no central star formation \citep[see also][]{2014arXiv1404.6926A}. The late accretion of high angular momentum gas
clearly favours the formation of extended galactic discs (see
\citealp{2013MNRAS.434.3142A}). 

\begin{figure}
\centering
	\includegraphics[width=0.4\textwidth]{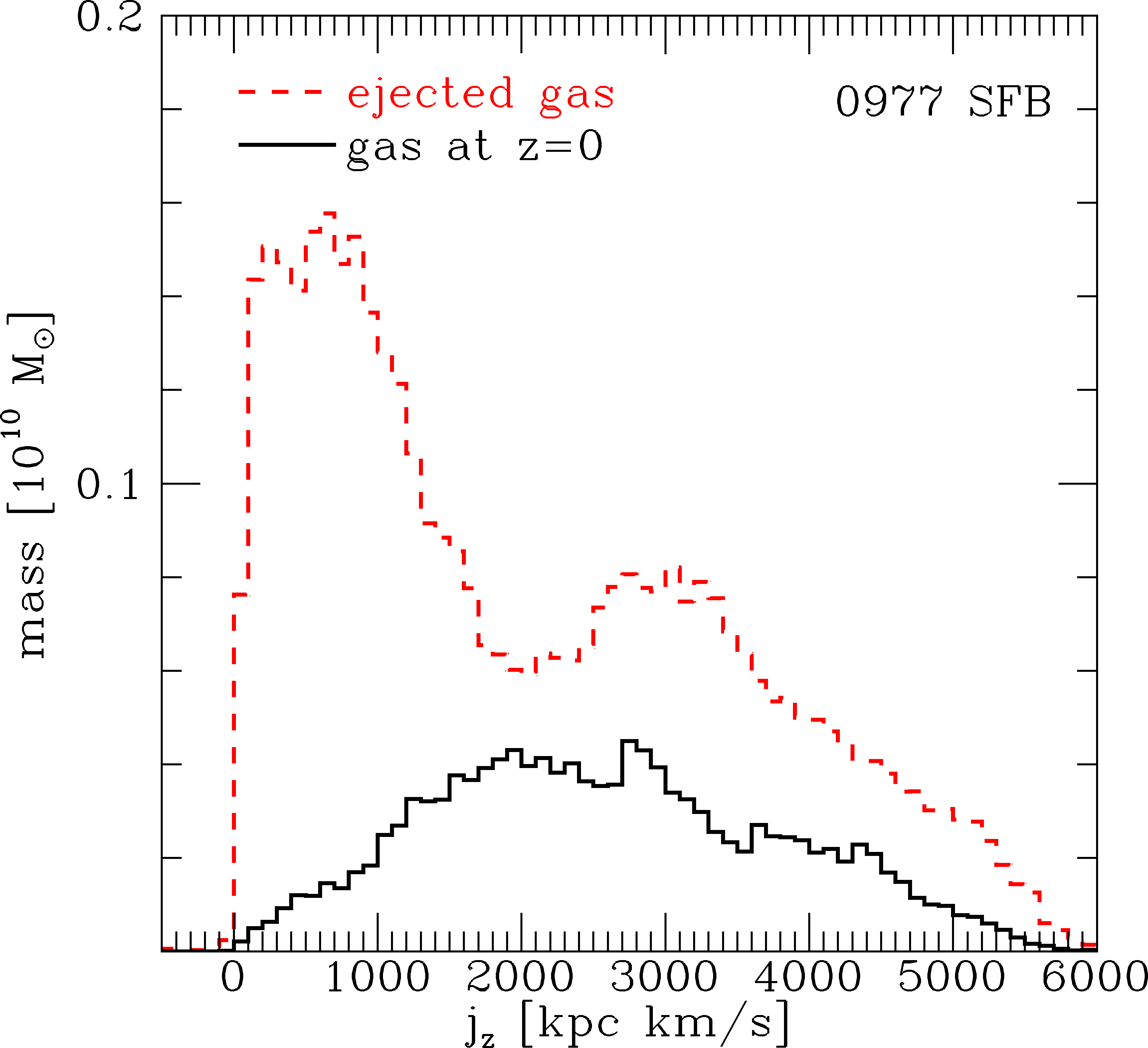}   
  \caption{The angular momentum distribution of disc gas particles at
    $z=0$ (black) and the angular momentum distribution for all particles that have been ejected from the disc from $z=4$ to the present day (red) for the SFB run of halo 0977. Ejected gas has predominantly low angular momentum as compared to the angular momentum of present-day disc gas.}
\label{jzout}
\end{figure}

We also record the $j_{z}$ value of a gas particle when the particle
leaves the disc. In the bottom row of Fig.~\ref{jej} we plot the $j_{z}$ values of gas particles which are ejected from the galactic disc by $z=0$ as a function of ejection time. In WFB particles are often ejected at late times, mostly caused by regular mergers (for halo 0977). This is in contrast to SFB, where gas is in general ejected at early times, also with low angular momentum, when the potential well is still shallow.
The predominant ejection of low angular momentum gas in the SFB model is also shown in Fig.~\ref{jzout}, where we plot a comparison of the angular momentum of expelled gas (red dashed) and of the angular momentum of the gas residing in the disc at $z=0$ (black) for the SFB model. 
Ejected gas has predominantly low angular momentum with a median of 1700 kpc km s$^{-1}$, while the average value for present-day disc gas is $\sim$2600 kpc km s$^{-1}$ (see also Fig.~\ref{jzout1646}, and Fig.~\ref{jzoutAqB} for an even more distinct case). This is in agreement with \citet{2011MNRAS.415.1051B} for SFB. 
(The enhanced ejection of high angular momentum gas during the last Gigayear in SFB is unique for halo 0977 and due to the ongoing merger at $z=0$ which warps the galactic disc.)

We can now determine whether cycling particles gain or lose angular momentum
during their journey through the galaxy. Most of the time the angular momentum of the gas
particles does not change significantly during their cycles
(Fig.~\ref{joutrec}). The ranges of $j_{z}$ of the cycling particles are very different with $j_{z,\mathrm{max}}\sim$ 3000 kpc km s$^{-1}$ in WFB and $j_{z,\mathrm{max}}\sim$ 5000 kpc km s$^{-1}$ in SFB. This underlines again one of the central effects of the strong feedback implementation, i.e. the enhanced (re-)accretion of gas at late times, that was already shown in Figs.~\ref{sfr} (top panel) and \ref{jej} (upper panels) and which is seen for all studied haloes. In WFB there is a weak trend
that some particles lose angular momentum. In SFB much more high
angular momentum gas is cycling and there is a trend that, in
particular, some of the gas ejected with low angular momentum gains
angular momentum during the cycling process. 

\begin{figure}
\centering
	\includegraphics[width=0.4\textwidth]{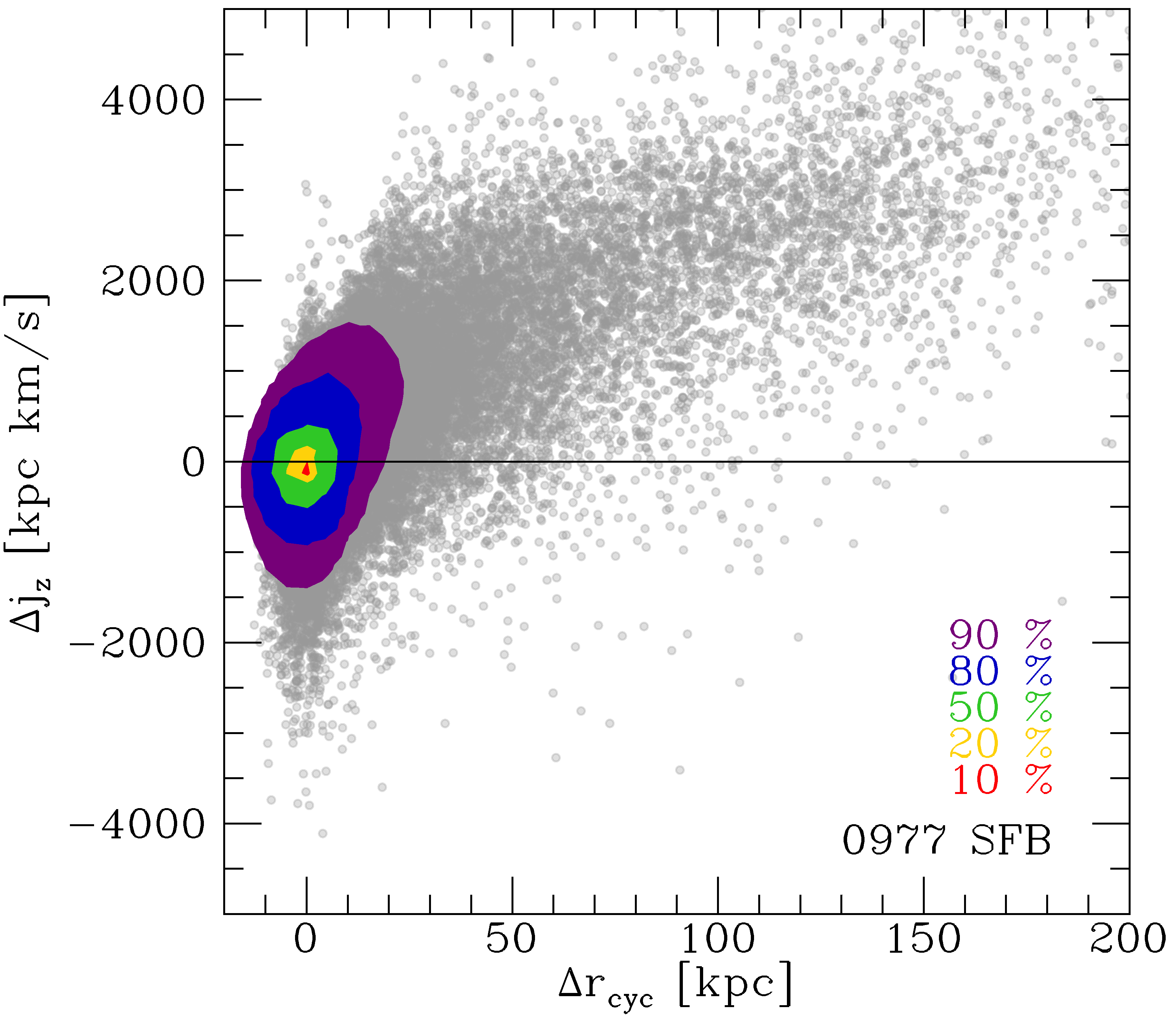} 
  \caption{Angular momentum change $\Delta j_{z}=j_{z\mathrm{,acc}} -
    j_{z\mathrm{,eject}}$ as a function of travel distance for the SFB run of halo 0977. There is a clear trend that
    particles expelled to large radii on
    average gain angular momentum.} 
\label{mdist_dj}
\end{figure}

If this gain in angular momentum originates from a direct influence of
the environment, e.g.\, the homogenisation of the angular momentum of cycling particles with that of the hot corona gas \citep{2012MNRAS.419..771B} or cosmic torques, we would expect that gas particles that travel far would gain more angular momentum. This is indeed true, as
shown in Fig.~\ref{mdist_dj}, where we plot the gain/loss of angular
momentum ($\Delta j_{\mathrm{z}}=j_{\mathrm{z,acc}} - j_{\mathrm{z,eject}}$) as a function
of travel distance for the SFB model. Particles moving further than $~\sim20$ kpc (for this halo) only gain
angular momentum. This is the case, for instance, for those
particles which are part of the bump in Fig.~\ref{tdelay} of
lately re-accreted particles with recycling times around 8 Gyr (also visible as the peak between 4 and 5.5 Gyr in Fig.~\ref{mdist_time}). All cycling gas particles
originating from this early merger gained angular momentum during the recycling process and appear in the upper
left part in the right panel of Fig.~\ref{joutrec}. The WFB
simulation is unaffected by this process as most particles travel
much less than 50 kpc. The trends towards the gain of angular momentum connected to travel distance in the SFB model are even stronger for the other haloes, especially for Aquarius B (cf. Fig.~\ref{mdist_djAqB}).

All the above processes (i) inefficient conversion into stars, (ii) high rates
of first accreted gas and delayed accretion of recycled gas
at high angular momentum, (iii) the mild angular momentum gain of gas
cycling at large radii and (iv) the predominant ejection of low angular
momentum gas, favour the formation of an extended disc in
the model with strong feedback.

\section{Summary}
\label{summary}

We present a comparative study of the gas assembly histories for five
cosmological galaxy zoom-in simulations
($\mathrm{M}_{\mathrm{vir}}=6.9\times10^{11} \mathrm{M}_{\odot}$ to
$1.7\times10^{12} \mathrm{M}_{\odot}$) each carried out twice, once with weak (WFB), and once with strong 
feedback (SFB) from massive stars. We summarize characteristic quantities of those simulations in Table~\ref{tab:4}.  The relative amount of stellar and
gas accretion at different redshifts and the angular momentum of the
accreted gas determine whether stars form and assemble in a
spheroid (low angular momentum) or a disc (high angular
momentum). We focus on the detailed 
evolutionary histories of the baryonic components since $z=4$ and track
accretion onto the halos and accretion and ejection from the
galaxies, as well as the full evolution of the disc gas angular momentum during
ejection and accretion.     

In the WFB simulations (see
e.g. \citealp{2010ApJ...725.2312O,2013arXiv1311.0284N}) the early 
($z>1$) conversion of low angular momentum gas into stars in the
galaxies and in the accreted substructures is favoured. This leads
to  the formation of systems with a high stellar-to-halo mass-ratio  
\citep{2010ApJ...725.2312O}, relatively low angular momentum, and a
significant spheroidal component
\citep{2012ApJ...754..115J,2013arXiv1311.0284N}. There is  
little late accretion of gas, which could form a (more extended)
disc-like component \citep[see also][]{2014arXiv1401.3180S}. Overall, the
behaviour reflects the well documented angular momentum
`problem' or `catastrophe'
(e.g. \citealp{2000ApJ...538..477N}).      
   
Simulations starting from identical initial conditions using a strong
feedback implementation show different behaviour and the
galaxies better resemble the observed population of present-day spiral galaxies 
\citep{2013MNRAS.434.3142A}. The early conversion of low angular
momentum gas into stars is -- in agreement with abundance matching estimates -- significantly suppressed. 
This leads to higher gas fractions at all redshifts and to flat star formation histories with less star formation at high redshift and more star formation at low redshift. 
This is a common feature of models with strong stellar feedback either directly coupled to the
surrounding ISM (see e.g \citealp{2010MNRAS.408..812S,2011MNRAS.410.2625P,2013MNRAS.434.3142A,2013MNRAS.428..129S})
or decoupled from the local ISM in a wind mode (see e.g \citealp{2008MNRAS.387..577O,2010MNRAS.406.2325O,2014MNRAS.437.1750M,2013MNRAS.436.2929H,2013MNRAS.436.3031V,2014MNRAS.tmp...38T}).
Our quantitative results regarding feedback efficiency and galaxy morphology agree with the qualitative conclusions for high-$z$ galaxies presented in \cite{2014ApJ...782...84A}.
Present-day gas is arranged in an extended disc. Most galactic stars in the SFB model form within the galaxy from disc gas, so the stellar component forms a disc and its angular momentum is a poor reflection of that of the dark matter halo. 

The gas accretion rates onto the central regions of the halo and the galactic disc are in general significantly higher in the SFB model (5-15 times). At early times ($z>1$) the accretion rates are dominated by first accreted gas, while the accretion of recycled gas becomes dominant at late times ($z<1$). The recycling of disc gas is an important feature of the SFB model with 50-60 per cent of the accreted gas mass participating in this process and thus forming a galactic fountain \citep{2008MNRAS.387..577O}.
Also, the high gas accretion rates at lower redshift decouple from the declining dark matter accretion \cite[see][]{2013MNRAS.436.2929H}.
In general, the angular momentum of first accreted as well as recycled gas increases significantly towards low redshift \cite[see also][]{2012MNRAS.419..771B}, resembling (for first accreted gas) an inside-out growth of the disc. This process might be influenced or delayed through mergers at early times.

A significant fraction (25-30 per cent) of the accreted disc gas is ejected and does not return by the present day. Confirming previous studies
\citep{2011MNRAS.415.1051B} this gas is predominantly ejected at high
redshift ($z>1$) - when the potential wells of the proto-galaxies are still
shallow - and has low angular momentum. The predominant ejection of
low angular momentum gas has long been proposed as a possible solution
to the angular momentum problem
\citep{2001MNRAS.321..471B,2002MNRAS.335..487M,2006MNRAS.372.1525D,2009MNRAS.396..141D}
and the promoting effect for the formation of discs has been confirmed
by cosmological simulations
\citep{2010Natur.463..203G,2011MNRAS.415.1051B}.
The ejected low angular momentum gas can by $z=0$ reach distances from the galaxy of the order of $\gtrsim$~1~Mpc and can thus contribute to metal enrichment of the IGM.

The efficient
ejection of low angular momentum gas also prevents conversion into stars even in major galaxy mergers. Mergers can trigger angular momentum loss of gas and have long been considered
a major problem for disc galaxy formation
\citep[e.g.][]{1996ApJ...471..115B}. Only recently has there been growing
evidence that - provided efficient star formation is suppressed and/or
the feedback is sufficiently strong - gas-rich early mergers are less
problematic
\citep{2005ApJ...622L...9S,2006ApJ...645..986R,2009ApJ...691.1168H,2009MNRAS.398..312G,2011MNRAS.415.3750M}
and are to some degree even required to explain the structural
evolution of the disc galaxy population (\citealp{2014arXiv1404.6926A}; see e.g.\, \citealp{2005A&A...430..115H} for observational evidence of disc reformation in the last 8 Gyrs after major mergers). 

The angular momentum of cycling gas particles is in general conserved, due to short typical travel distances (less than 10 kpc for 60-85 per cent of the gas) and short recycling times ($<$~1~Gyr).
Some smaller fraction ($\sim$10 per cent) of cycling gas -
mostly ejected during early turbulent phases of merging - gains more than 1000 kpc km s$^{-1}$ before its re-accretion, eventually through mixing with the hot corona gas \citep{2012MNRAS.419..771B} or from cosmic torques. 
However, if gas particles leave the disc for long times ($>$~1~Gyr) and travel large distances ($\gtrsim$ 50-100 kpc), then they always gain angular momentum before re-accretion. 
However, this gas contributes little ($<$ 3 per cent) to the total gas accretion. 
Nonetheless, the general trend towards larger travel distances and longer recycling times in SFB as compared to WFB is strong.

All the above processes resulting from strong stellar feedback favour the formation of extended galactic discs.

\section*{Acknowledgments}
We thank the anonymous referee for valuable comments on the draft.
TN and MA acknowledge support from the DFG excellence cluster `Origin and Structure of the Universe'.

\bibliographystyle{mn2e}
\bibliography{./references.bib}

\clearpage
\appendix

\section{Halo 1192} 
\label{1192}

\begin{figure*}
\centering
	\includegraphics[width=0.78\textwidth]{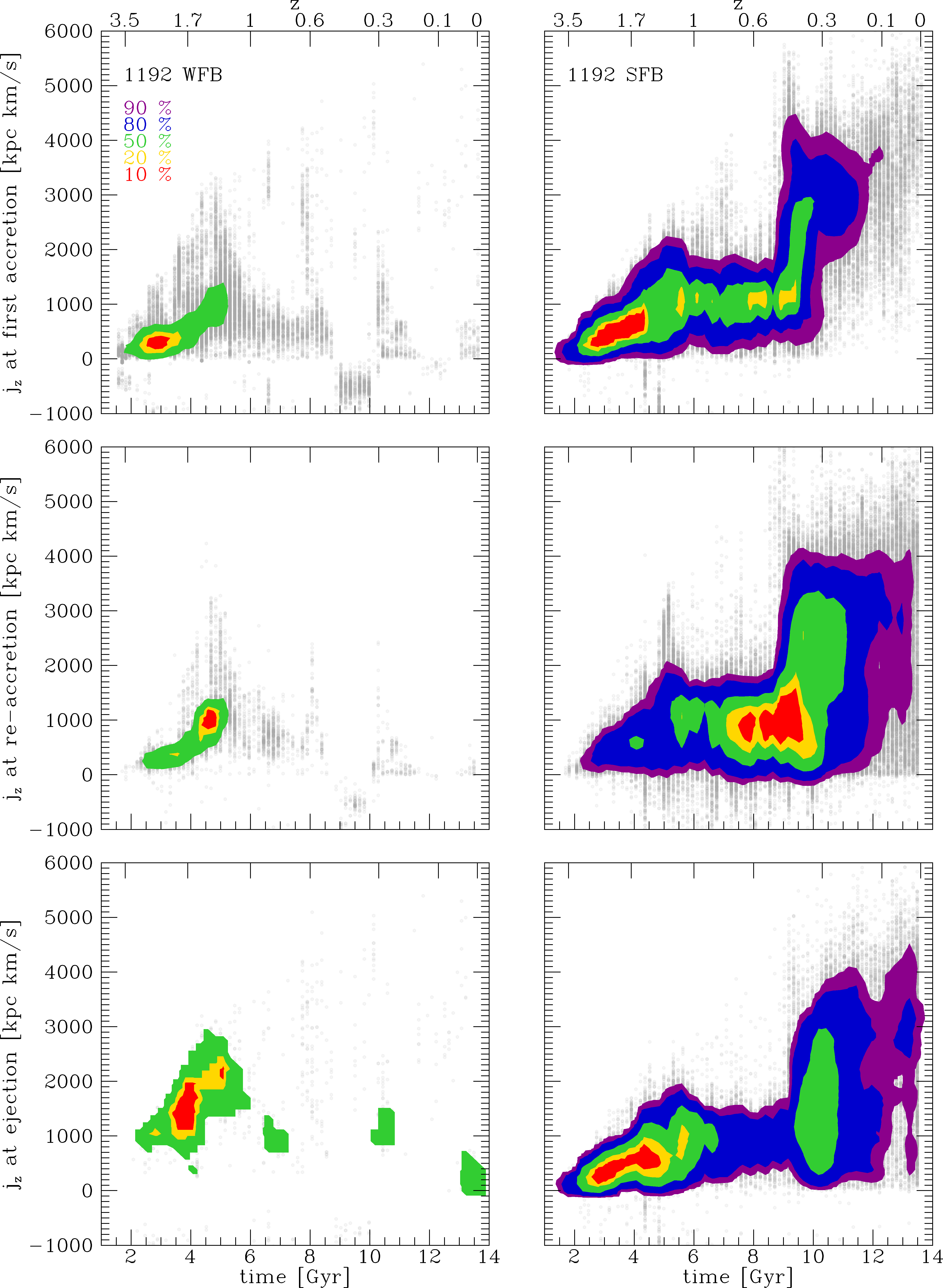}
  \caption{Compare Fig.~\ref{jej}. Top panel: Angular
          momentum in $z$-direction of first-time accreted particles as a
          function of accretion time for WFB (left panel) and SFB
          (right panel) in the 1192 models. Caused by the merger activity until $\sim$ 10 Gyr in the SFB model the gas particles are accreted with low angular momentum \citep[see][]{2014arXiv1404.6926A}. For the WFB run, the impact of halo interaction and misaligned gas accretion can be seen after $z\sim1$.
          Middle panel: Same as top panel but for re-accreted particles as
    a function of accretion time for WFB (left panel) and SFB (right
    panel) (cf. middle panel of Fig.~\ref{jej}). During the merger period, star formation is concentrated towards the center and gas particles are ejected and re-accreted with relatively low angular momentum in SFB.
    	Bottom panel: Same as top panel but now for particles that are ejected from the disc by $z=0$ as a function of the time of ejection for WFB (left panel) and SFB (right panel).}  
\label{jej1192}
\end{figure*}

\noindent\begin{minipage}{0.48\textwidth}
The evolution of halo 1192 with the strong feedback model is characterized by several gas rich minor mergers, especially between $z=1.1$ and $z=0.4$, which have great influence on some galaxy properties we want to show. The travel distances are on average higher (cf. Fig.~\ref{mdist_time1192}) as compared to halo 0977, leading to a higher fraction of angular momentum gain (cf. Fig.~\ref{joutrec1192}), while the early inside-out growth of the galaxy is prevented by the merger period (cf. top panel in Fig.~\ref{jej1192}). In the weak feedback model, there is some halo interaction and misaligned gas accretion at late times.
\vspace{1cm}
\end{minipage}

\noindent\begin{minipage}{0.98\textwidth}
\centering
	\includegraphics[width=0.78\textwidth]{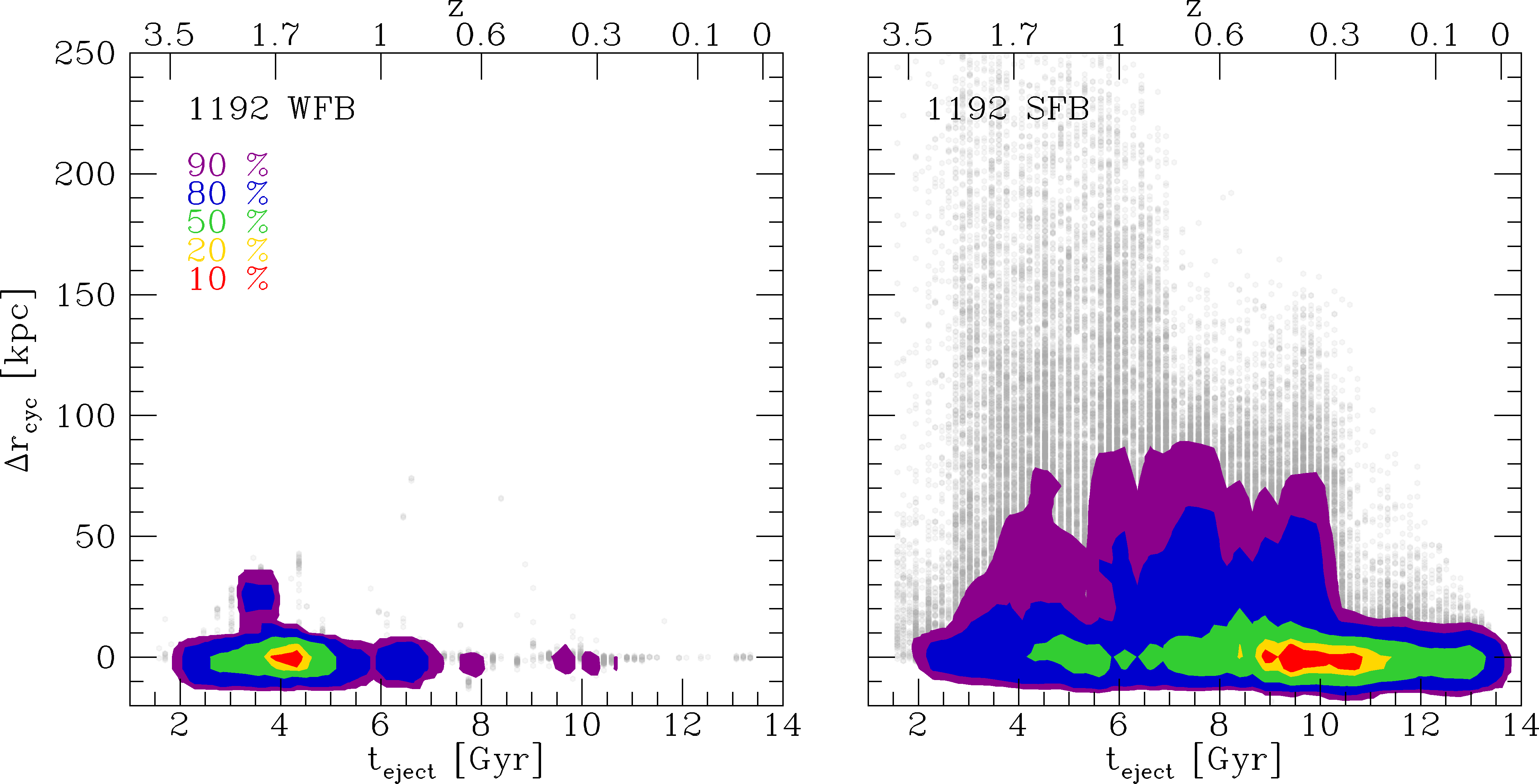} 
  \captionof{figure}{Compare Fig.~\ref{mdist_time}. Travel distance as a function of ejection time for WFB (left panel) and SFB (right panel) in the 1192 models. Individual particles are shown as grey dots,
          the coloured regions indicate the respective fractions at the
          indicated levels. In SFB there are continuous gas rich minor mergers that cause the ejection of gas to relatively large radii ($\gtrsim$ 50 kpc).}\label{mdist_time1192}
\end{minipage}

\noindent\begin{minipage}{0.98\textwidth}
\centering
	\includegraphics[width=0.78\textwidth]{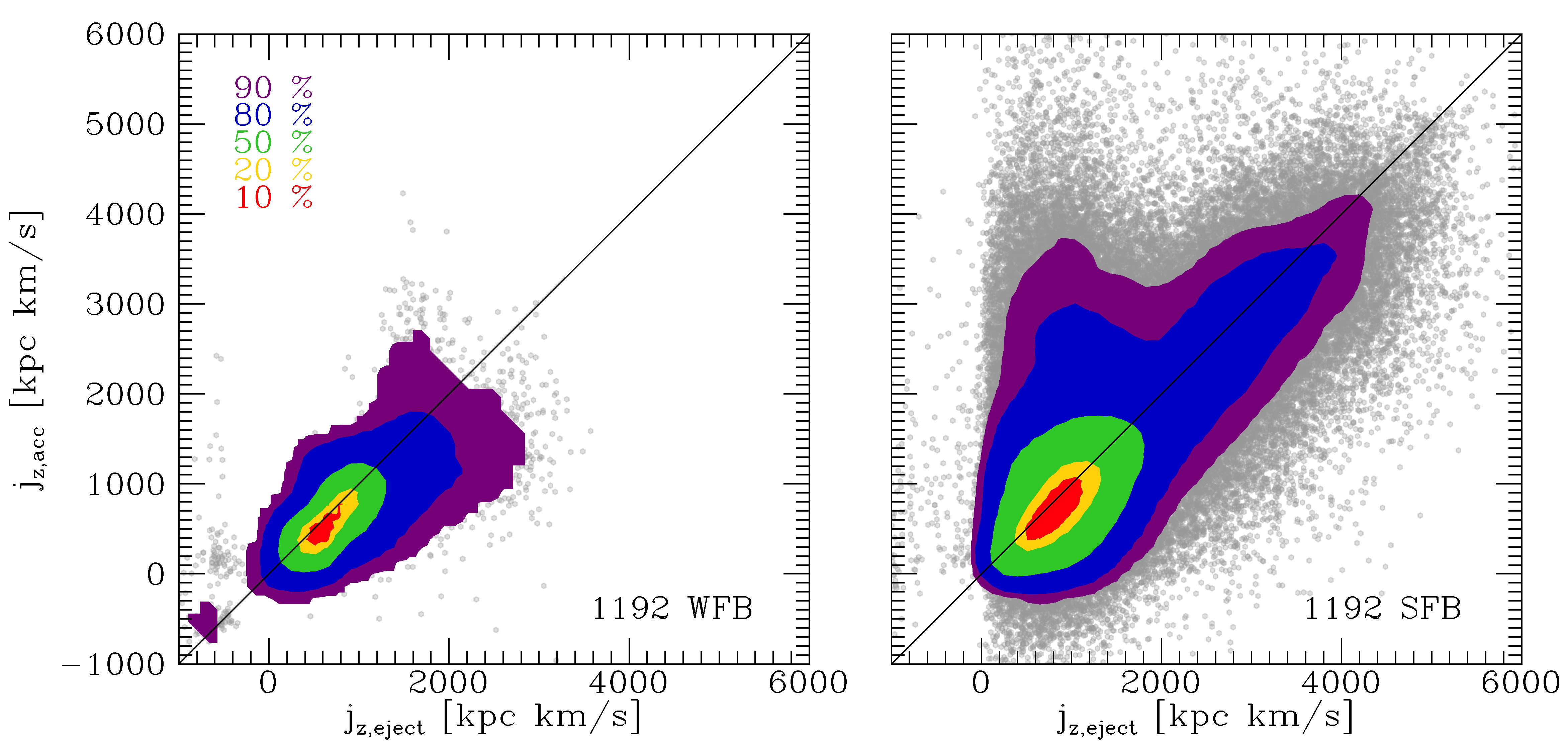} 
   \captionof{figure}{Compare Fig.~\ref{joutrec}. Angular momentum at ejection $j_{\mathrm{z,eject}}$
     vs. angular momentum at accretion $j_{\mathrm{z,acc}}$ of 
     cycling gas particles for WFB (left panel) and SFB (right
     panel) in the 1192 models. The connection between on average longer travel distances (cf. Fig.~\ref{mdist_time1192}) and more angular momentum gain is visible.}  
\label{joutrec1192}
\end{minipage}

\noindent\begin{minipage}{0.48\textwidth}
		\centering
		\vspace{0.5cm}
		\begin{tabular}{lrrrr} 
		\hline
		mass $[10^{10} \mbox{M}_{\odot}]$ & WFB & SFB \\ \hline
		\textbf{total accreted stars} & \textbf{9.02} & \textbf{0.70}\\
		\textbf{total accreted gas} & \textbf{6.78} & \textbf{22.29}\\
		\textbf{first accreted gas} & \textbf{5.97} & \textbf{13.08}\\
		\hspace{2mm}\textbullet\hspace{2mm} $\%$ cycling & 1 & 5 \\
		\hspace{2mm}\textbullet\hspace{2mm} $\%$ not cycling & 4 & 4 \\
		\hspace{2mm}\textbullet\hspace{2mm} $\%$ condensed within r$_{15}$ & 59 & 49\\
		\hspace{2mm}\textbullet\hspace{2mm} $\%$ ejected by $z=0$ & 36 & 42 \\
		\hline
		\end{tabular}
		\captionof{table}{Respective masses of the different accretion modes onto r$_{15}$ for halo 1192 as shown for halo 0977 in Fig.~\ref{accmodes} and Table~\ref{tab:3}. The amount of condensed star mass in SFB is higher compared to halo 0977.} 
		\label{tab:31192}
    	\label{acc1192}
\end{minipage}

\clearpage

\section{Halo 1646} 
\label{1646}

\begin{figure*}
\vspace{0.5cm}
\centering
	\includegraphics[width=0.78\textwidth]{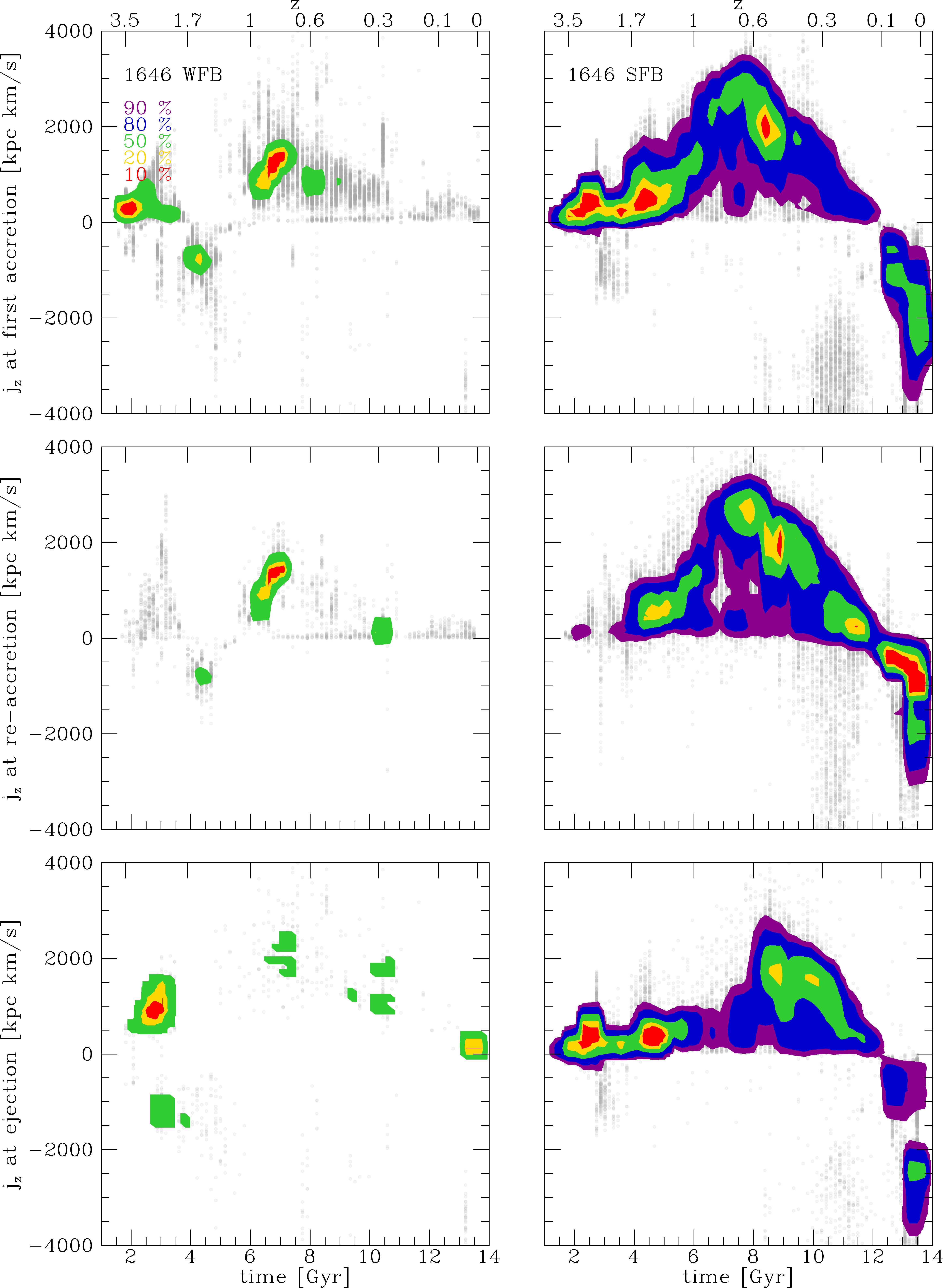}
  \caption{Compare Fig~\ref{jej}. Top panel: Angular
          momentum in $z$-direction of first-time accreted particles as a
          function of accretion time for WFB (left panel) and SFB
          (right panel) in the 1646 models. There is a drastic change in the angular momentum orientation of the accreted gas in SFB starting from $\sim$ 7 Gyr, which is to some extent also visible in WFB. For SFB, it leads to pure counter-rotating infall after 12 Gyrs (compare discussion in \citealt{2013MNRAS.434.3142A}; 2014).
  	Middle panel: Same as top panel but for re-accreted particles as
    a function of accretion time for WFB (left panel) and SFB (right
    panel) (cf. middle panel of Fig.~\ref{jej}). The angular momentum direction of re-accreted gas changes corresponding to that of first accreted gas.
  	Bottom panel: Same as top panel but now for particles that are ejected from the disc by $z=0$ as a function of the time of ejection for WFB (left panel) and SFB (right panel). Also for gas that is ejected by $z=0$ is the re-orientation of the angular momentum vector visible.}  
\label{jej1646}
\end{figure*}

\noindent\begin{minipage}{0.48\textwidth}
The assembly history of halo 1646 is particularly interesting as it experiences a dramatic change in the angular momentum orientation of infalling gas starting at $z\sim0.7$ (cf. Fig.~\ref{jej1646}), leading in the SFB model to the formation of a young counter-rotating stellar disc that lives on top of the older stellar disc (for more details, see \citealp{2013MNRAS.434.3142A}; 2014), while the gaseous disc flips completely. Subsequent accretion of gas onto the central regions of the galaxy triggers star formation and with it ejection of gas up to large radii (cf. Fig.~\ref{mdist_time1646}), giving cause to above average re-accretion of particles with recycling times between 1 and 5 Gyrs within the last Gyr (cf. Fig.~\ref{tdelay1646} and middle panel of Fig.~\ref{jej1646}). There is significant angular momentum gain of cycling particles belonging to the old co-rotating disc, as well as significant angular momentum loss of those particles that are re-accreted after $z\sim0.7$ (cf. Figs.~\ref{joutrec1646} and \ref{mdist_dj1646}). The galactic evolution is reflected in the differing angular momentum distribution of present-day and by $z=0$ ejected galactic disc gas particles (cf. Fig.~\ref{jzout1646}).
\end{minipage}

\noindent\begin{minipage}{0.98\textwidth}
\centering
	\includegraphics[width=0.78\textwidth]{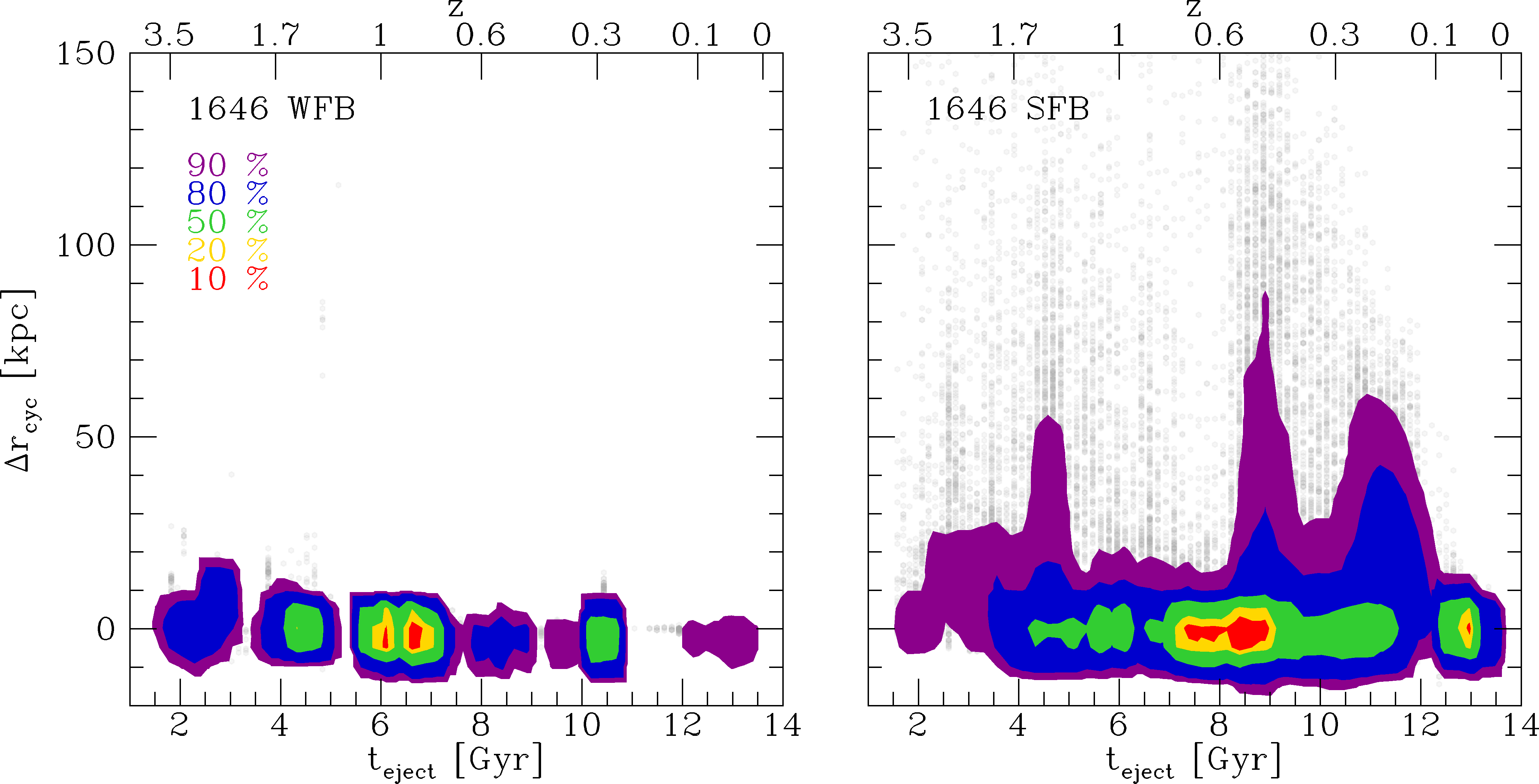} 
  	\captionof{figure}{Compare Fig.~\ref{mdist_time}. Travel distance as a function of ejection time for WFB (left panel) and SFB (right panel) in the 1646 models. The peak at $\sim4.5$ Gyr in the SFB model (right panel) is caused by a merger event. There are no mergers at 9 and 11 Gyrs but accretion of gas particles to the central regions (cf. top panel of Fig.~\ref{jej1646}) which triggers star formation.} 
\label{mdist_time1646}
\end{minipage}

\noindent\begin{minipage}{0.98\textwidth}
		\centering
		\includegraphics[width=0.5\textwidth]{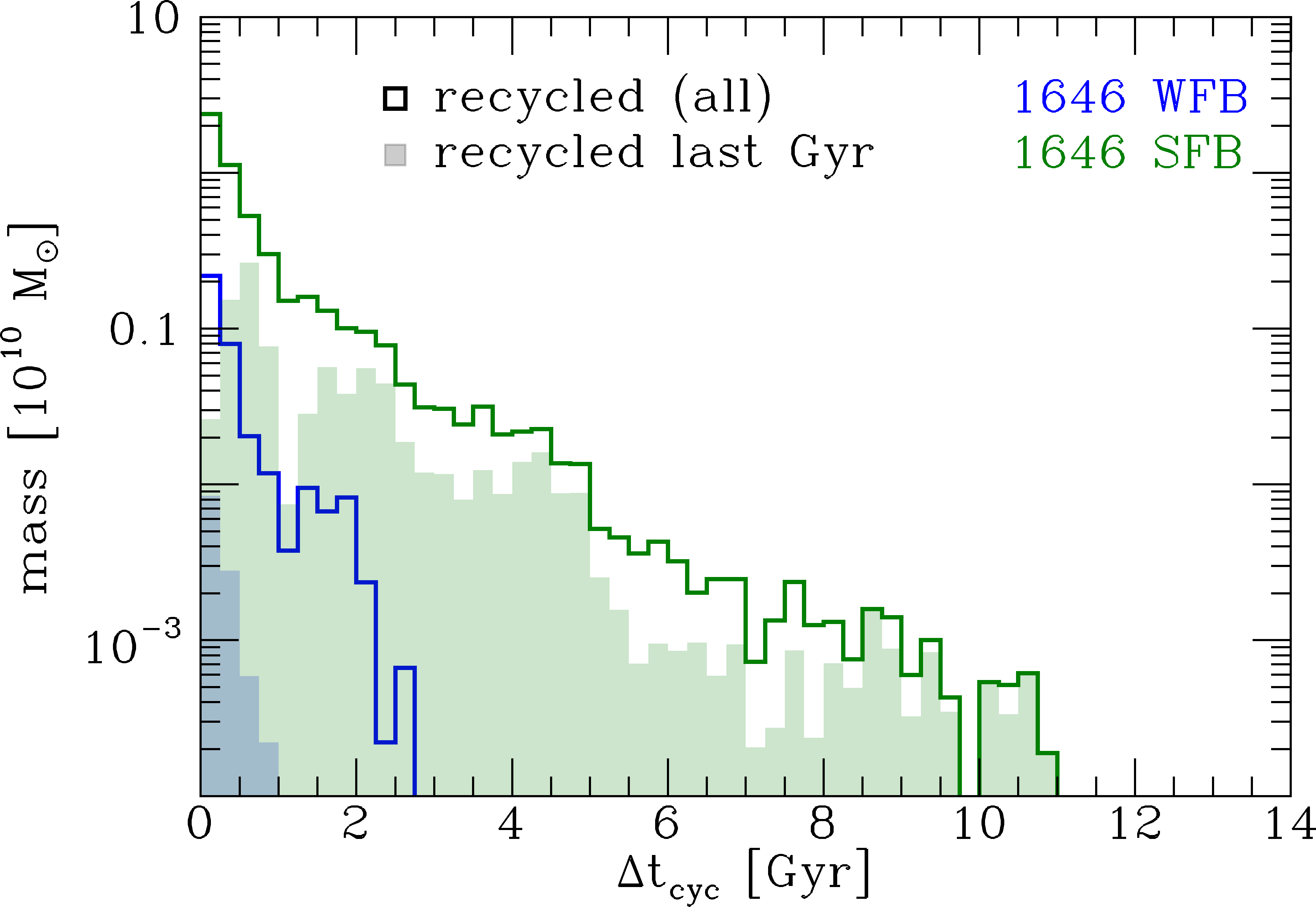}  
  		\captionof{figure}{Compare Fig.~\ref{tdelay}. Recycling times for WFB (blue) and SFB (green) in Gigayears in the 1646 models. Particle mass which is re-accreted within the last Gigayear is shown as the filled areas. The fraction of particles with 1 Gyr $< \Delta\mathrm{t}_{\mathrm{cyc}} <$ 5 Gyrs is very high, corresponding to the ejection of particles between 8 and 12 Gyrs (cf. Fig.~\ref{mdist_time1646}).}
		\label{tdelay1646}
\end{minipage}

\noindent\begin{minipage}{0.48\textwidth}
		\centering
		\vspace{2cm}
		\begin{tabular}{lrrrr} 
		\hline
		mass $[10^{10} \mbox{M}_{\odot}]$ & WFB & SFB \\ \hline
		\textbf{total accreted stars} & \textbf{5.99} & \textbf{0.41}\\
		\textbf{total accreted gas} & \textbf{6.01} & \textbf{11.23}\\
		\textbf{first accreted gas} & \textbf{5.05} & \textbf{7.15}\\
		\hspace{2mm}\textbullet\hspace{2mm} $\%$ cycling & 1 & 12 \\
		\hspace{2mm}\textbullet\hspace{2mm} $\%$ not cycling & 5 & 6 \\
		\hspace{2mm}\textbullet\hspace{2mm} $\%$ condensed within r$_{15}$ & 51 & 29\\
		\hspace{2mm}\textbullet\hspace{2mm} $\%$ ejected at $z=0$ & 43 & 53 \\
		\hline
		\end{tabular}
	\captionof{table}{Respective masses of the different accretion modes onto r$_{15}$ for halo 1646 as shown for halo 0977 in Fig.~\ref{accmodes} and Table~\ref{tab:3}.}
		\label{tab:31646}
\end{minipage}

\clearpage

\clearpage

\noindent\begin{minipage}{0.98\textwidth}
\centering
	\includegraphics[width=0.78\textwidth]{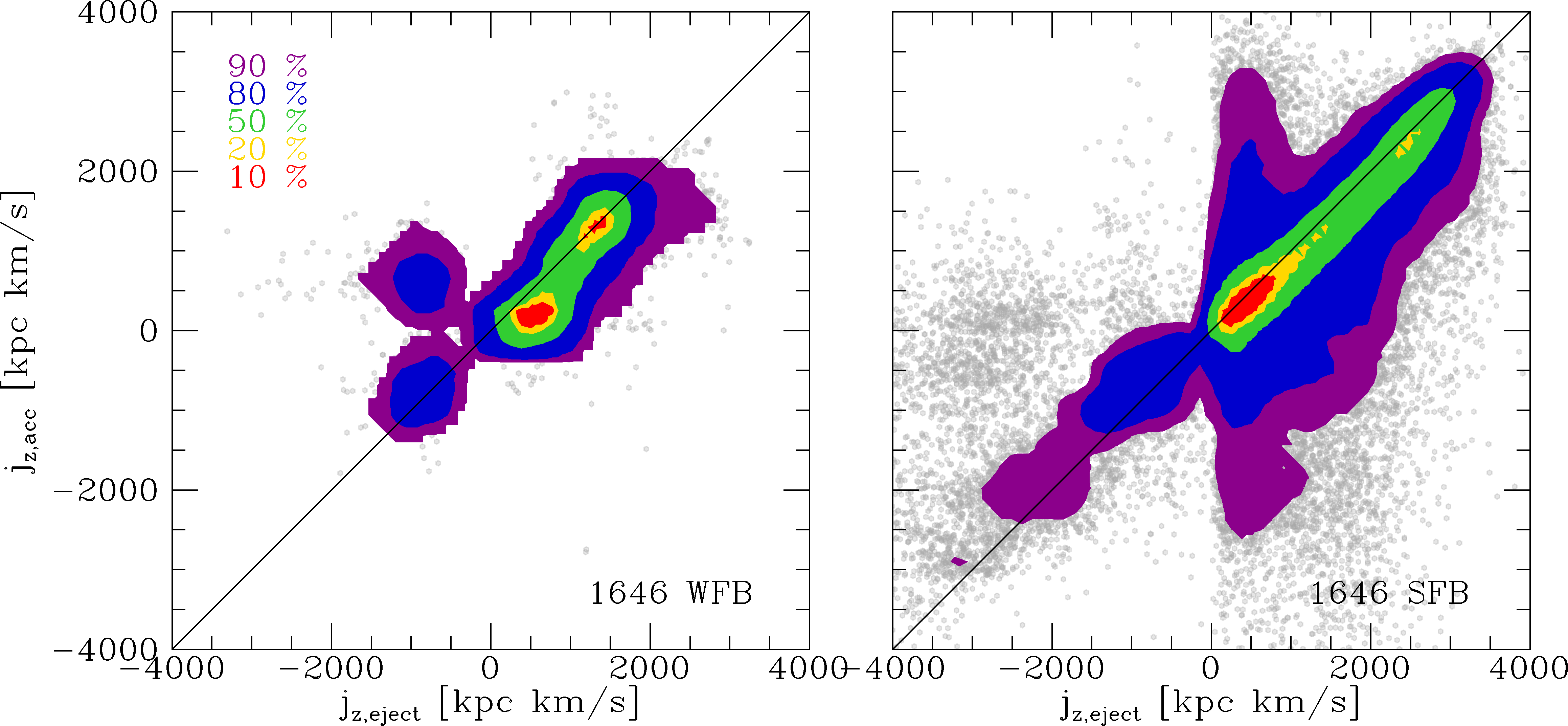} 
   \captionof{figure}{Compare Fig. \ref{joutrec}. Angular momentum at ejection $j_{\mathrm{z,eject}}$
     vs. angular momentum at accretion $j_{\mathrm{z,acc}}$ of 
     cycling gas particles for WFB (left panel) and SFB (right
     panel) in the 1646 models. For both models the re-orientation of the angular momentum of (re-)accreted gas is distinguishable.}  
\label{joutrec1646}
\end{minipage}

\noindent\begin{minipage}{0.98\textwidth}
\centering
	\includegraphics[width=0.4\textwidth,]{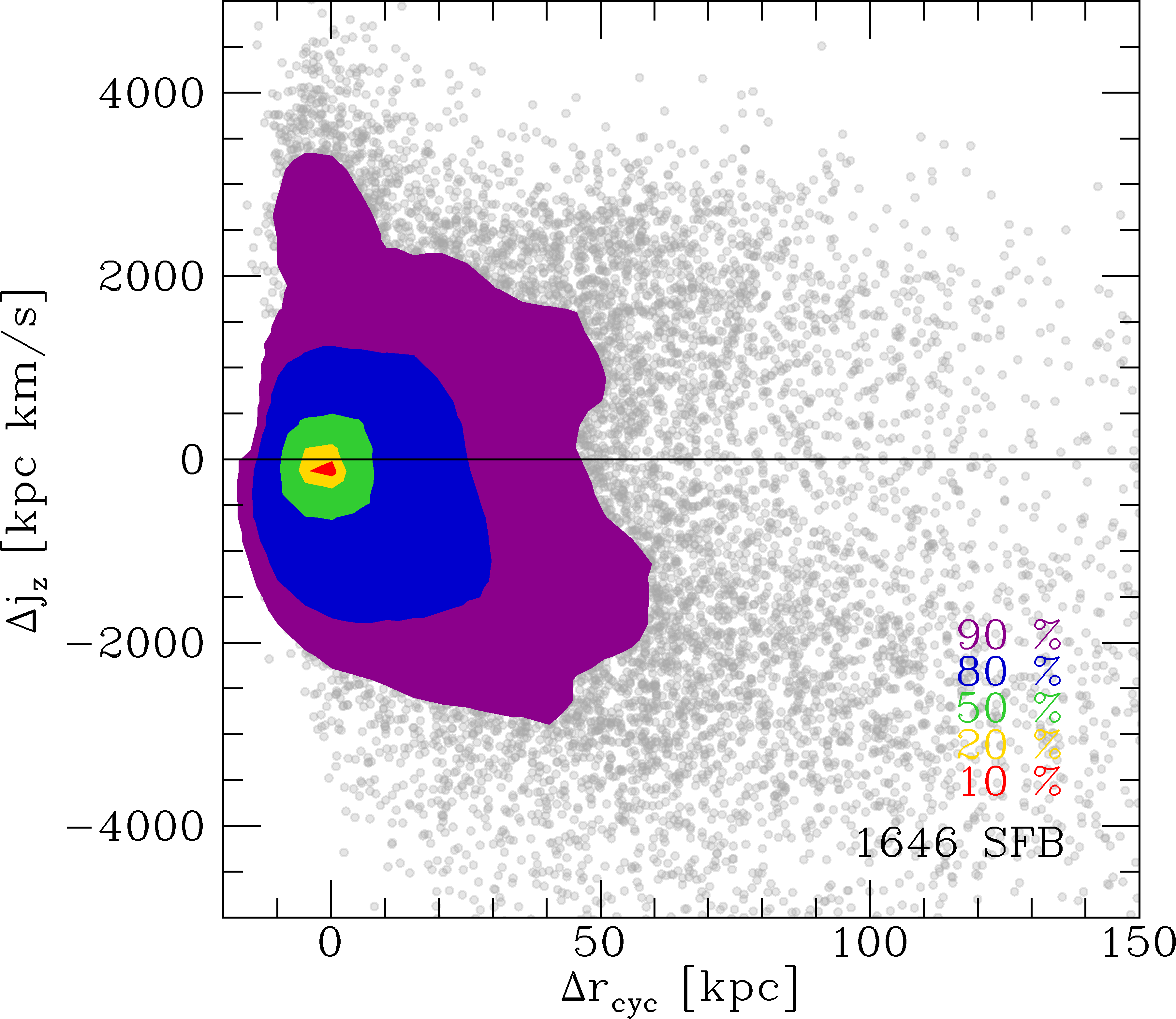} 
  \captionof{figure}{Compare Fig. \ref{mdist_dj}. Angular momentum change $\Delta j_{\mathrm{z}}=j_{\mathrm{z,acc}} -
    j_{\mathrm{z,eject}}$ as a function of travel distance for the SFB run of model 1646. Through the re-orientation of the angular momentum of the cycling gas particles in SFB the distribution of the data points is mirrored on the horizontal axis in contrast to the other haloes: Long travel distances are correlated with either angular momentum gain (before $z\sim0.7$) or loss (after $z\sim0.7$) with respect to the original orientation of the gas disc.} 
\label{mdist_dj1646}
	\includegraphics[width=0.4\textwidth]{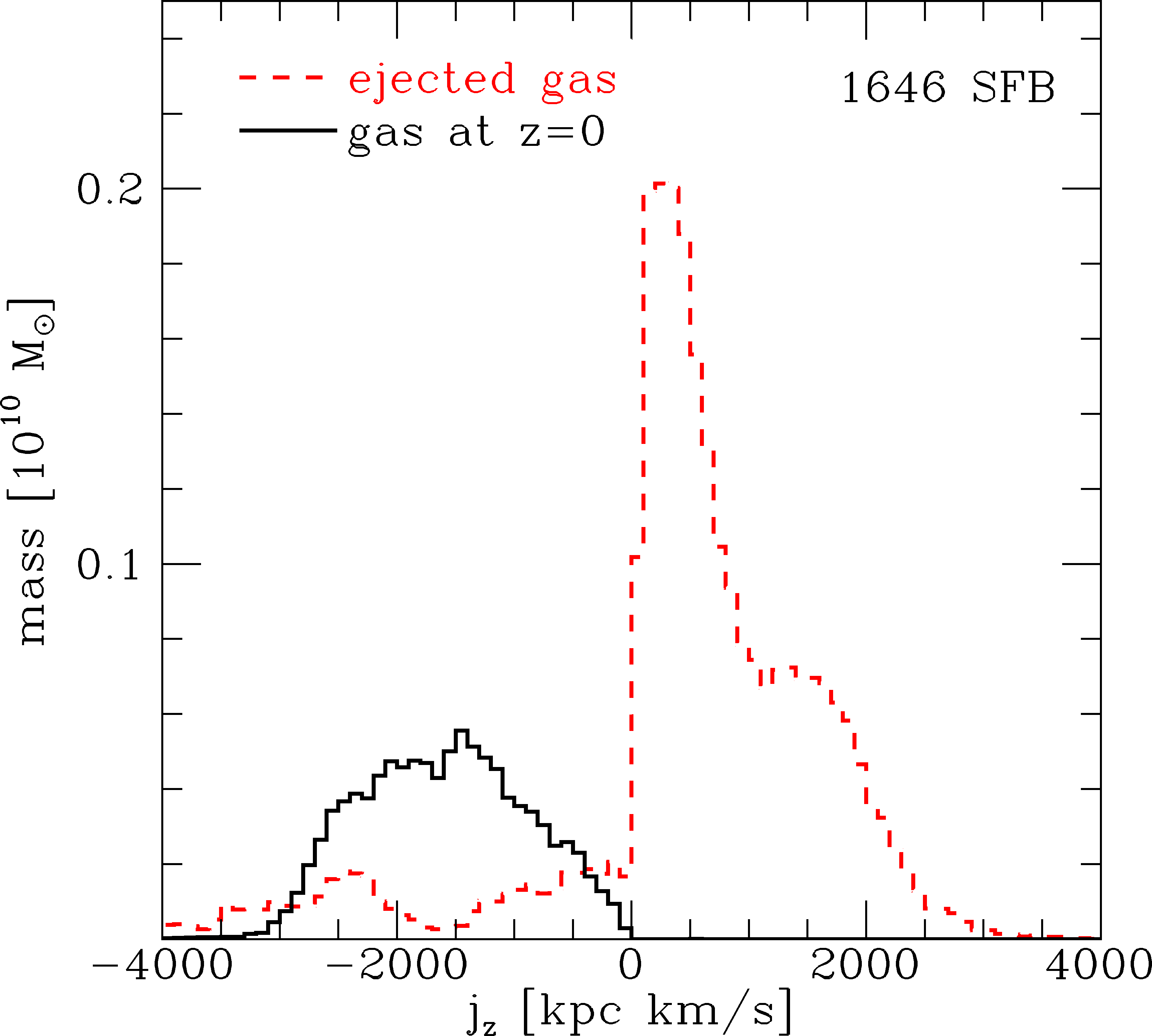}   
  \captionof{figure}{Compare Fig.~\ref{jzout}. Angular momentum distribution of disc gas particles at
    $z=0$ (black) and of finally ejected particles (red; at ejection)
    for the SFB run of model 1646. The re-orientation of the gas disc is obvious, where the present-day gas disc is counter-rotating, contrary to most of the ejected gas in previous times.}
\label{jzout1646}
\end{minipage}

\clearpage

\section{Halo Aquarius B} 
\label{AqB}

Halo Aquarius B experiences a major merger at $z\sim1.1$ in the SFB model after which starts the inside-out growth of the galaxy (not shown here). The merger event is reflected in a distinct peak of angular momentum gain of initially low angular momentum gas particles (cf. Figs.~\ref{joutrecAqB} and \ref{mdist_djAqB}). The ejection of low angular momentum gas particles primarily during the early (merger) phase of the SFB galaxy results in a particular angular momentum distribution of present-day and at $z=0$ ejected disc gas particles (cf. Fig.~\ref{jzoutAqB}). For this halo, we also add a temperature diagnostic to show that there is no evidence for cold blobs in the strong feedback model (cf. Fig.~\ref{TdistAqB} and Section~\ref{simulations}).

\noindent\begin{minipage}{0.98\textwidth}
\vspace{1.5cm}
\centering
	\includegraphics[width=0.78\textwidth]{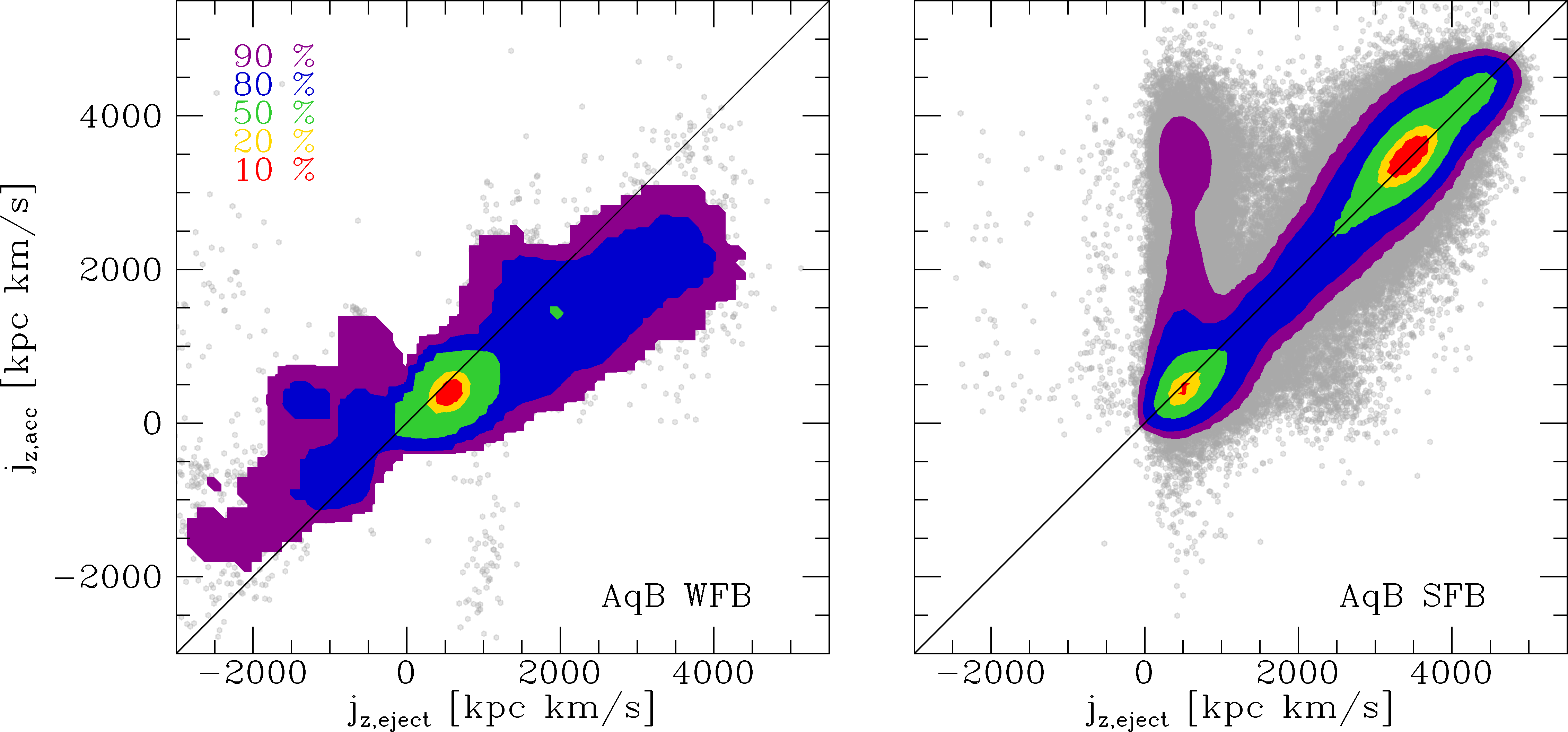} 
   \captionof{figure}{Compare Fig.~\ref{joutrec}. Angular momentum at ejection $j_{\mathrm{z,eject}}$
     vs. angular momentum at accretion $j_{\mathrm{z,acc}}$ of 
     cycling gas particles for WFB (left panel) and SFB (right
     panel) in the AqB models. For WFB we can see the signature of counter-rotation gas particles. There is also a trend towards angular momentum loss. Opposed to that, there is a clear trend towards angular momentum gain of low angular momentum gas particles during the recycling process in SFB, connected to long travel distances of gas particles involved in a major merger.}  
\label{joutrecAqB}
\end{minipage}

\noindent\begin{minipage}{0.98\textwidth}
\centering
	\includegraphics[width=0.4\textwidth]{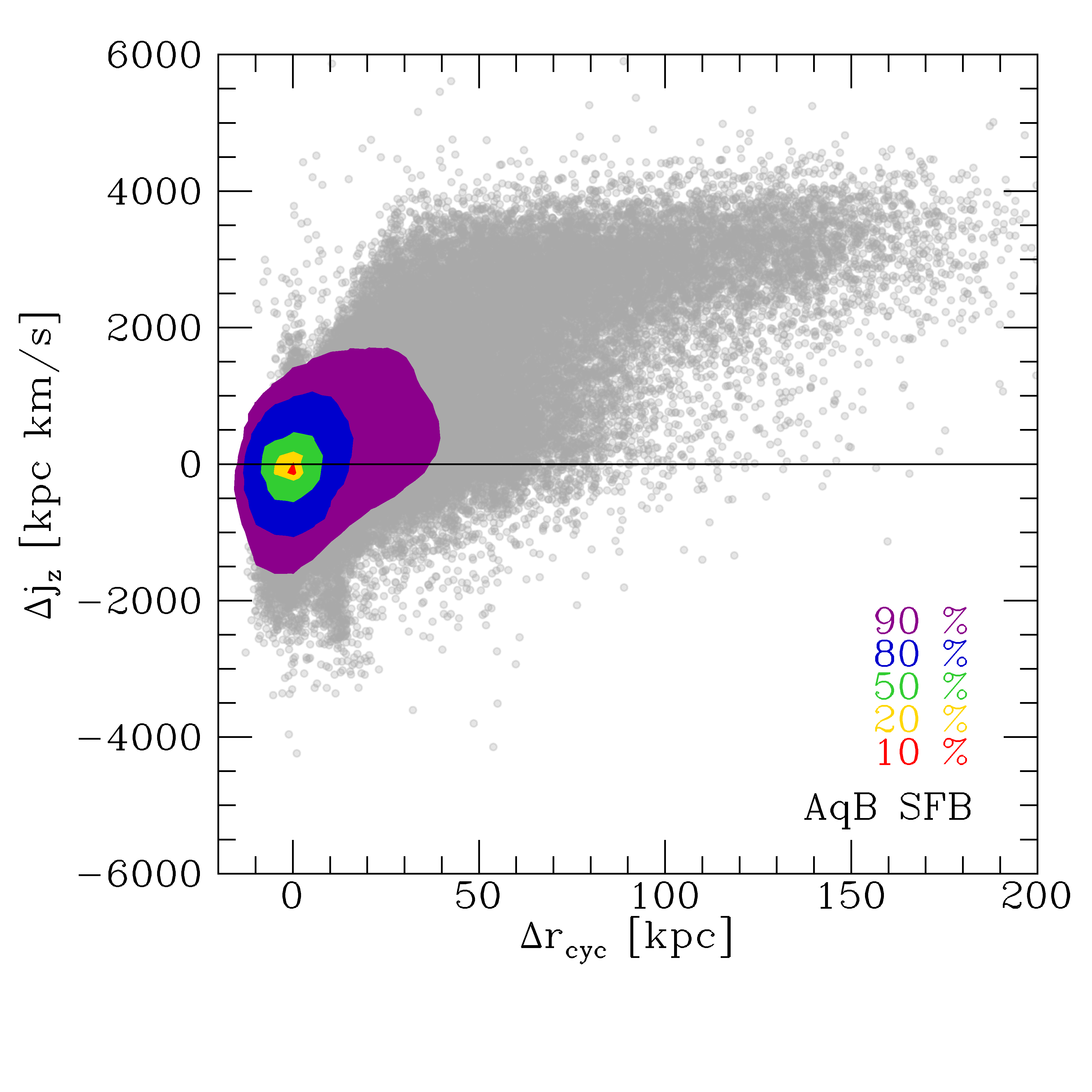} 
  \captionof{figure}{Compare Fig. \ref{mdist_dj}. Angular momentum change $\Delta j_{\mathrm{z}}=j_{\mathrm{z,acc}} -
    j_{\mathrm{z,eject}}$ as a function of travel distance for the SFB run of model AqB. The trend towards angular momentum gain for particles in SFB with travel distances above a certain threshold as described in the main text (cf. Section~\ref{trav}) is particularly clear in this figure.} 
\label{mdist_djAqB}
\end{minipage}

\noindent\begin{minipage}{0.48\textwidth}
		\centering
		\vspace{0.5cm}
		\begin{tabular}{lrrrr} 
		\hline
		mass $[10^{10} \mbox{M}_{\odot}]$ & WFB & SFB \\ \hline
		\textbf{total accreted stars} & \textbf{6.35} & \textbf{0.65}\\
		\textbf{total accreted gas} & \textbf{3.19} & \textbf{14.15}\\
		\textbf{first accreted gas} & \textbf{2.59} & \textbf{7.03}\\
		\hspace{2mm}\textbullet\hspace{2mm} $\%$ cycling & 2 & 11 \\
		\hspace{2mm}\textbullet\hspace{2mm} $\%$ not cycling & 6 & 4 \\
		\hspace{2mm}\textbullet\hspace{2mm} $\%$ condensed within r$_{15}$ & 52 & 33\\
		\hspace{2mm}\textbullet\hspace{2mm} $\%$ ejected by $z=0$ & 40 & 52 \\
		\hline
		\end{tabular}
		\label{tab:3AqB}
	\captionof{table}{Respective masses of the different accretion modes onto r$_{15}$ for halo AqB as shown for halo 0977 in Fig.~\ref{accmodes} and Table~\ref{tab:3}.}
	\vspace{10.2cm}
\end{minipage}

\clearpage

\noindent\begin{minipage}{0.98\textwidth}
\centering
	\includegraphics[width=0.4\textwidth]{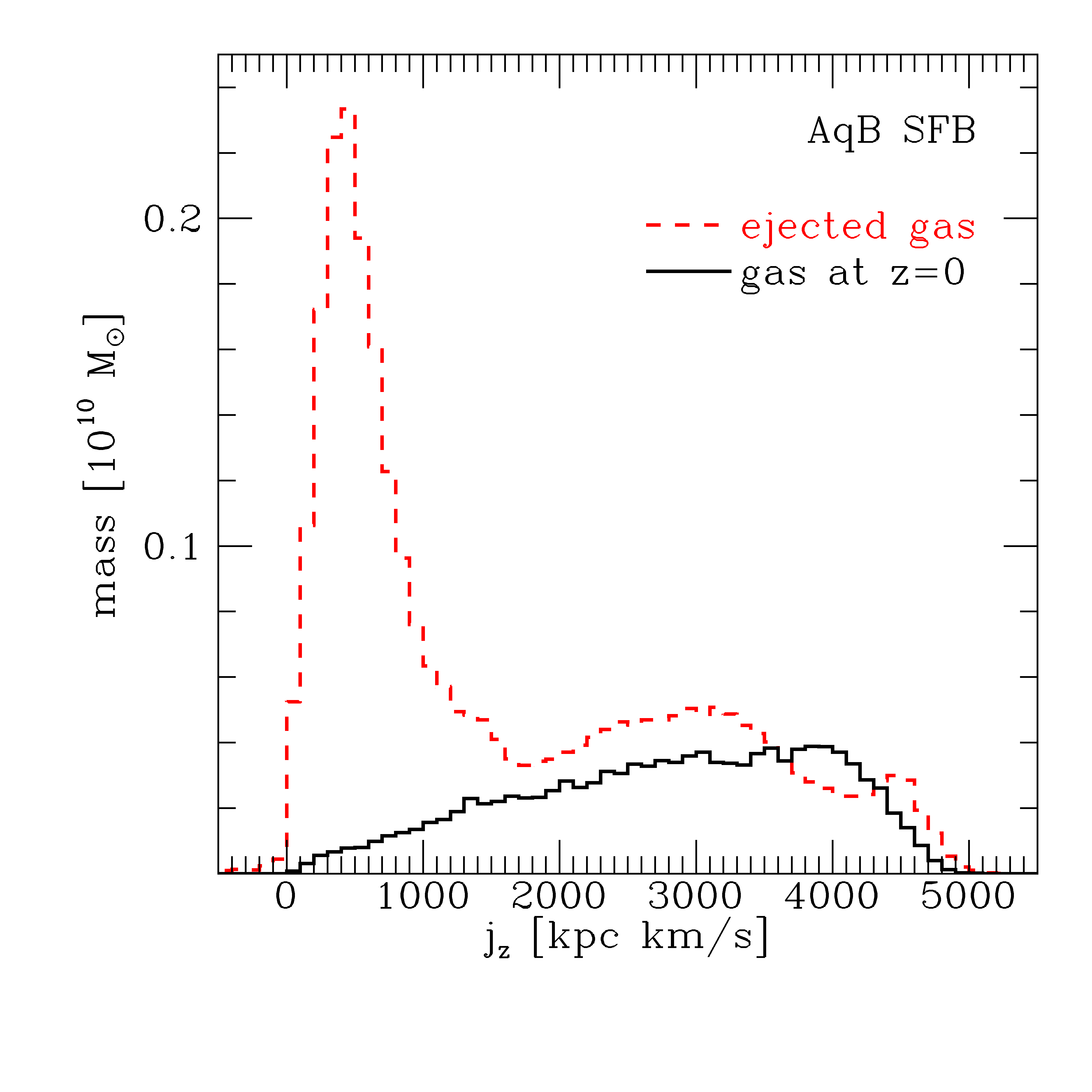}   
  \captionof{figure}{Compare Fig. \ref{jzout}. Angular momentum distribution of disc gas particles at
    $z=0$ (black) and of finally ejected particles (red; at ejection)
    for the SFB run of model AqB.}
\label{jzoutAqB}
	\includegraphics[width=0.4\textwidth]{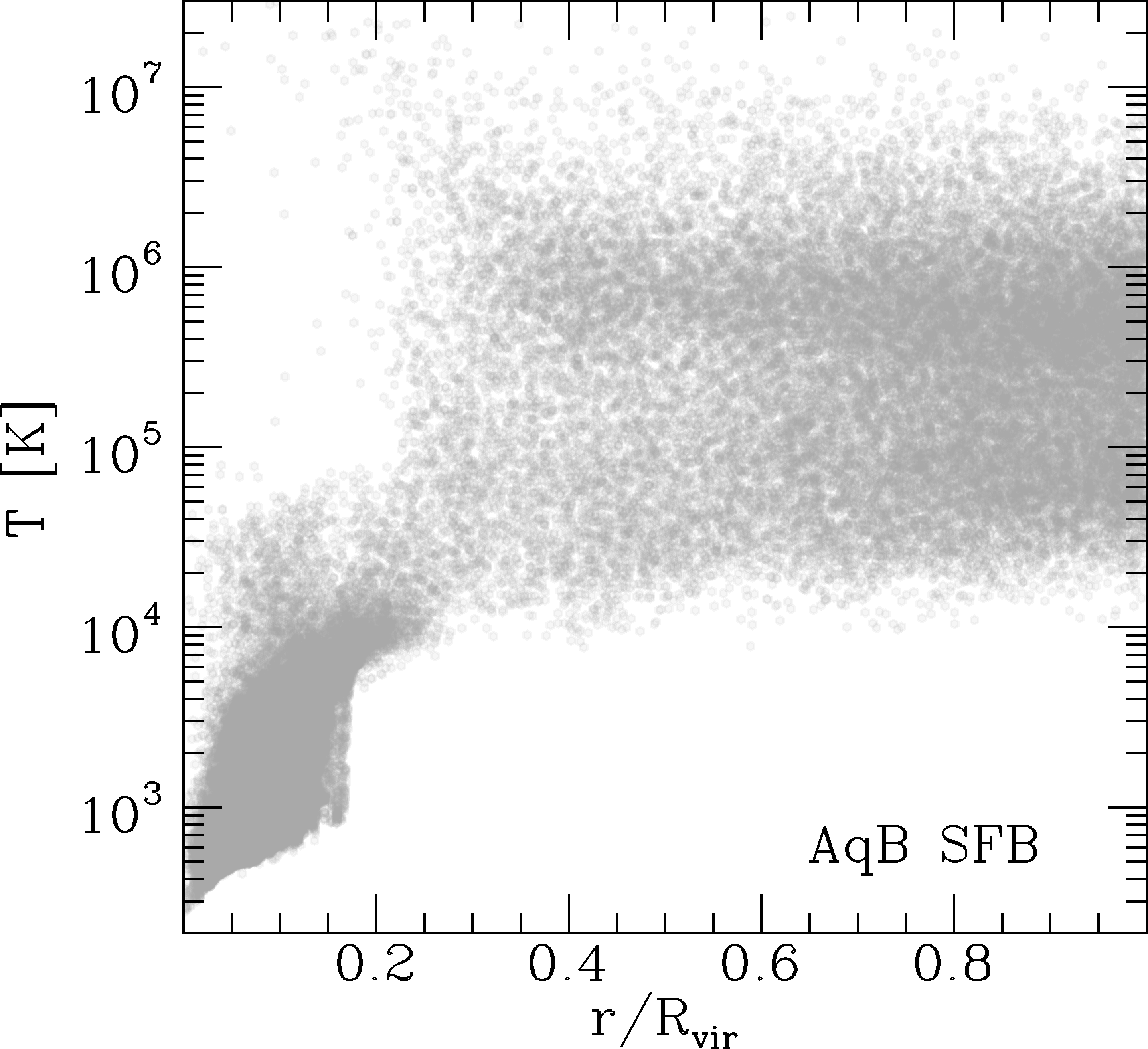}   
  \captionof{figure}{Temperature of gas particles within the halo as a function of distance from the galactic center at $z=0$. The cold dense gas disc is constituted of particles with $T<10^{4}$K and $r/\mathrm{R}_{\mathrm{vir}}\lesssim0.15$. Particles with $T>10^{4}$K and $r/\mathrm{R}_{\mathrm{vir}}>0.2$ are counted among diffuse halo gas particles. There is no evidence for cold gas clumps which would be located roughly in the region of $T<10^{4}$K and $r/\mathrm{R}_{\mathrm{vir}}>0.2$. See Section~\ref{simulations} in the main text for a discussion.}
\label{TdistAqB}
\end{minipage}

\clearpage

\section{Halo Aquarius D} 
\label{AqD}

Halo Aquarius D is special in the presented set of haloes because of its quiescent evolution after $z=0.4$ and its low accretion rates at late times. This can be seen in Figs.~\ref{gasrateAqD} and \ref{tdelayAqD}. Note that irrespective of that, the re-accretion of particles dominates over first accretion since $z=1$. There is much star formation (cf. Table~\ref{tab:3AqD}) especially before $z=0.4$ which causes relatively high average travel distances of cycling gas particles (cf. filled green area in Fig.~\ref{gasrateAqD}).
\vspace{2cm}

\noindent\begin{minipage}{0.48\textwidth}
	\centering
		\includegraphics[width=0.94\textwidth]{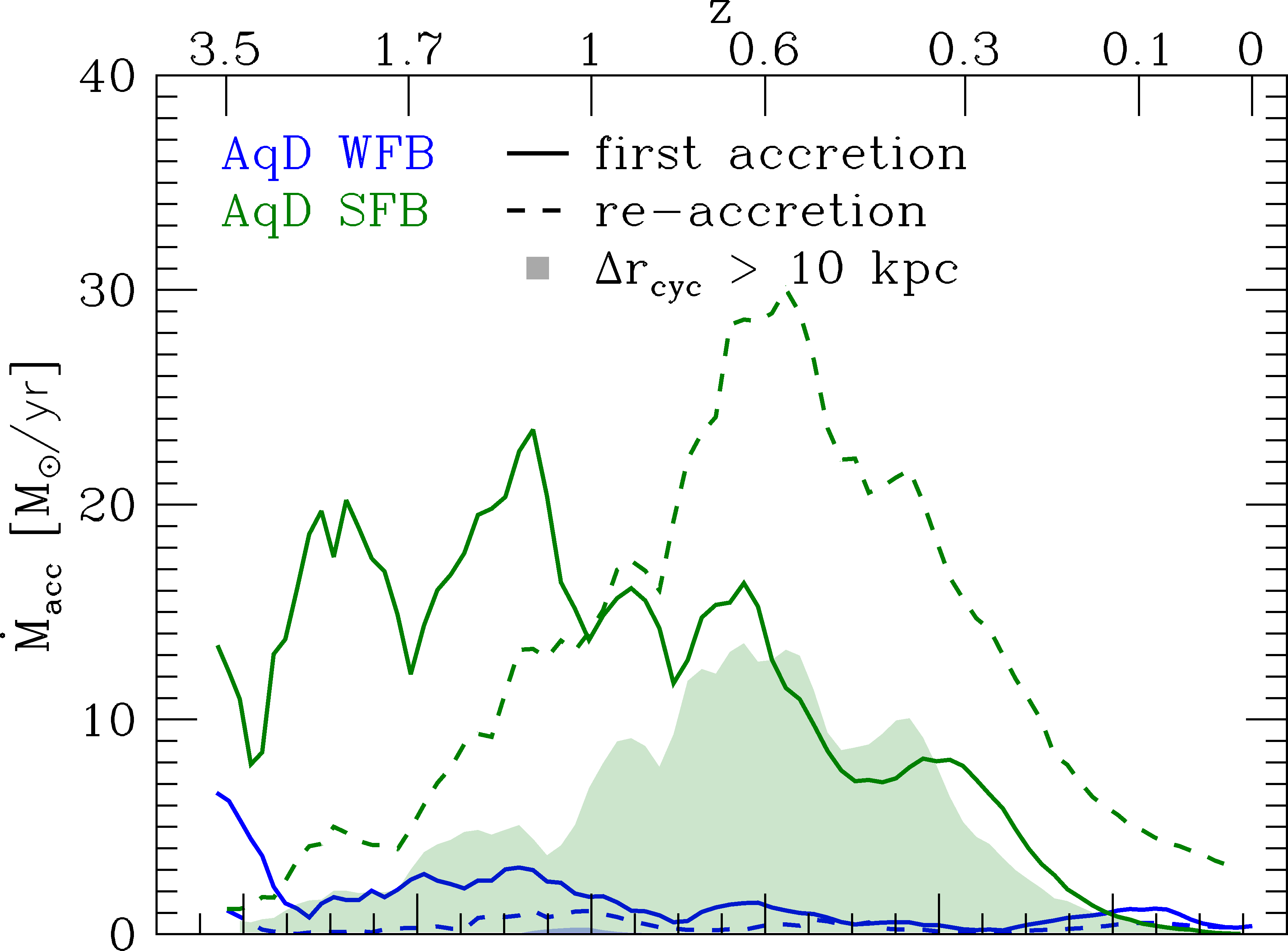} 
 		 \captionof{figure}{Compare top panel of Fig. \ref{sfr}. Gas accretion rate onto the disc in the AqD models separated into first
    accretion (solid lines) and re-accretion (dashed lines; filled
    regions for particles with $\Delta \mathrm{r}_{\mathrm{cyc}} >$ 10
    kpc) as a function of cosmic time and redshift for WFB (blue) and SFB
    (green).}  
		\label{gasrateAqD}
\end{minipage}

\noindent\begin{minipage}{0.98\textwidth}
\vspace{8cm}
\end{minipage}

\noindent\begin{minipage}{0.48\textwidth}
		\centering
		\vspace{1cm}
		\begin{tabular}{lrrrr} 
		\hline
		mass $[10^{10} \mbox{M}_{\odot}]$ & WFB & SFB \\ \hline
		\textbf{total accreted stars} & \textbf{11.99} & \textbf{1.10}\\
		\textbf{total accreted gas} & \textbf{7.48} & \textbf{25.81}\\
		\textbf{first accreted gas} & \textbf{6.25} & \textbf{16.23}\\
		\hspace{2mm}\textbullet\hspace{2mm} $\%$ cycling & 2 & 2 \\
		\hspace{2mm}\textbullet\hspace{2mm} $\%$ not cycling & 4 & 2 \\
		\hspace{2mm}\textbullet\hspace{2mm} $\%$ condensed within r$_{15}$ & 46 & 45\\
		\hspace{2mm}\textbullet\hspace{2mm} $\%$ ejected by $z=0$ & 48 & 52 \\
		\hline
		\end{tabular}
		\captionof{table}{Respective masses of the different accretion modes onto r$_{15}$ for halo AqD as shown for halo 0977 in Fig.~\ref{accmodes} and Table~\ref{tab:3}. The amount of condensed stellar mass in SFB is higher as compared to halo 0977.} 
		\label{tab:3AqD}
\end{minipage}

\vspace{15.5mm}
\noindent\begin{minipage}{0.48\textwidth}
		\centering
		\includegraphics[width=0.99\textwidth]{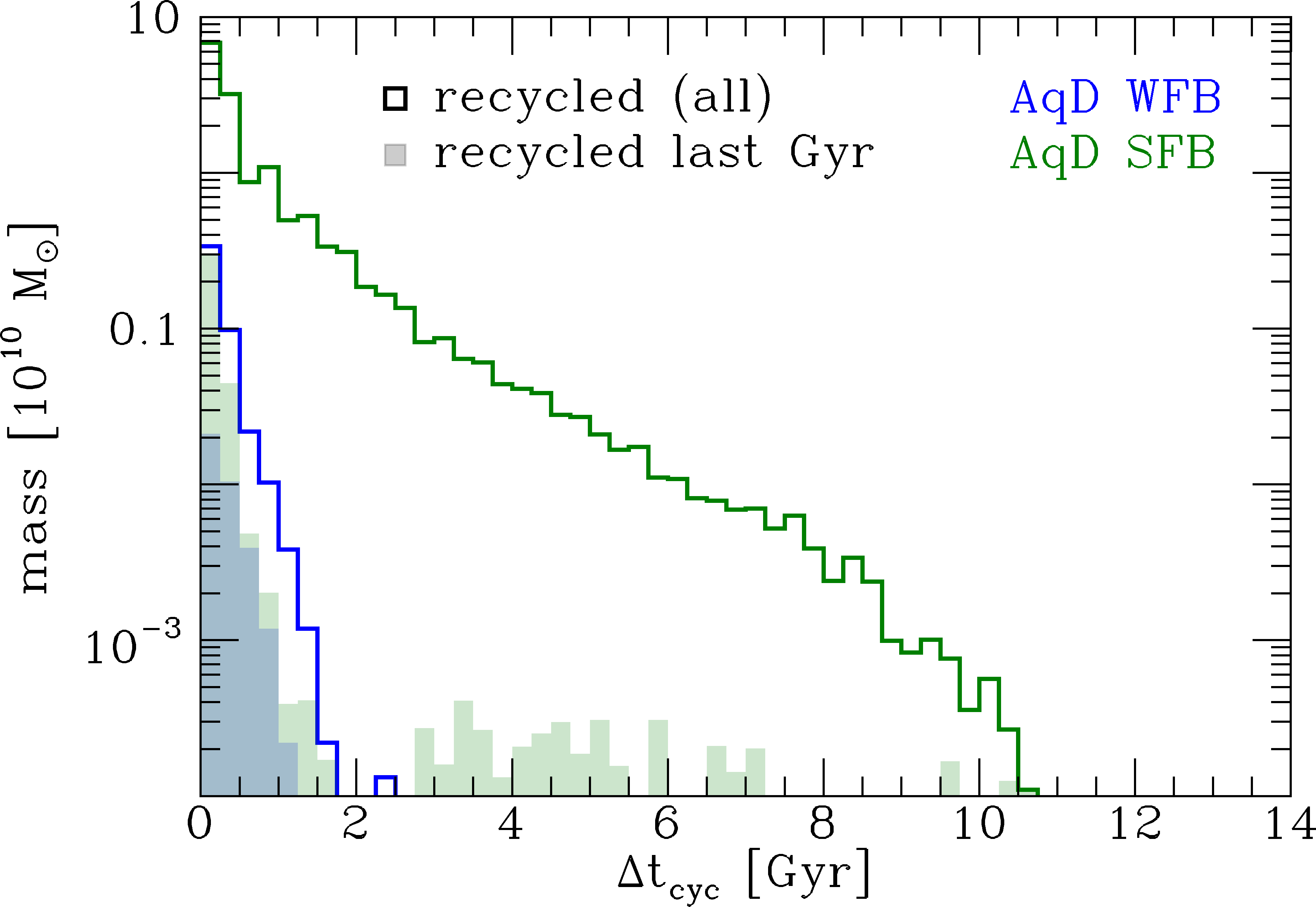} 
 		 \captionof{figure}{Compare Fig.~\ref{tdelay}. Travel distance as a function of ejection time for WFB (left panel) and SFB (right panel) in the AqD models. As can be also seen in Fig.~\ref{gasrateAqD}, there is little late re-accretion in SFB (green filled area).}  
		\label{tdelayAqD}
\end{minipage}

\label{lastpage}
\end{document}